\documentclass[english,prb,twocolumn,showpacs, superscriptaddress,floatfix,longbibliography]{revtex4-1}

\usepackage[T1]{fontenc}
\usepackage[utf8]{inputenc}
\usepackage{amsmath,amssymb,amstext}
\usepackage{xcolor}
\usepackage{graphicx}
\usepackage{esint}
\usepackage{babel}
\usepackage{verbatim}
\usepackage{color}
\usepackage{subfigure}
\usepackage{braket}
\usepackage[pdfborder={0 0 1}]{hyperref}
\definecolor{FSB-color}{named}{magenta}

\definecolor{MR-color}{named}{teal}

\begin{document}

\title{Spectral properties of Andreev crystals}

\author{Mikel Rouco}
\email{mikel.rouco@ehu.eus}
\address{Centro de F\'isica de Materiales (CFM-MPC), Centro Mixto CSIC-UPV/EHU, Manuel de Lardizabal 5, E-20018 San Sebasti\'an, Spain}

\author{F Sebastian Bergeret}
\email{fs.bergeret@csic.es}
\address{Centro de F\'isica de Materiales (CFM-MPC), Centro Mixto CSIC-UPV/EHU, Manuel de Lardizabal 5, E-20018 San Sebasti\'an, Spain}
\address{Donostia International Physics Center (DIPC), Manuel de Lardizabal 4, E-20018 San Sebastian, Spain}

\author{Ilya V Tokatly}
\email{ilya.tokatly@ehu.es}
\address{Donostia International Physics Center (DIPC), Manuel de Lardizabal 4, E-20018 San Sebastian, Spain}
\address{Nano-Bio Spectroscopy group and European Theoretical Spectroscopy Facility (ETSF), Departamento de Pol\'imeros y Materiales Avanzados: F\'isica, Qu\'imica y Tecnolog\'ia, Universidad del
  Pa\'is Vasco, Av. Tolosa 72, E-20018 San Sebasti\'an, Spain}
\address{IKERBASQUE, Basque Foundation for Science, E48009 Bilbao, Spain}

\date{\today}

\begin{abstract}
 We present an exhaustive study of Andreev crystals (ACs) -- quasi-one-dimensional superconducting wires with a periodic distribution of magnetic regions. The exchange field in these regions is assumed to be much smaller than the Fermi energy. Hence, the transport through the magnetic region can be described within the quasiclassical approximation. In the first part of the paper, by assuming that the separation between the magnetic regions is larger than the coherence length, we derive the effective   nearest-neighbour tight-binding equations for ACs with a helical magnetic configuration. The spectrum within the gap of the host superconductor shows a pair of energy-symmetric bands. By increasing the strength of the magnetic impurities in ferromagnetic ACs, these bands cross without interacting. However,  in any other helical configuration,  there is a value of the magnetic strength at which the bands touch each other, forming a Dirac point. Further increase  of the magnetic strength leads to a system with an inverted gap.   We study junctions between  ACs with inverted spectrum and show that 
 junctions between (anti)ferromagnetic ACs (always) never exhibit bound states at  the interface. 
 In the second part,  we extend our analysis beyond the  nearest-neighbour approximation by solving the Eilenberger equation for infinite  and junctions between semi-infinite ACs with collinear magnetization. From the obtained  quasiclassical Green functions,  we compute the local density of states and the local spin polarization  in anti- and ferromagnetic ACs.  We show that these junctions may exhibit bound states at the interface and fractionalization of the surface spin polarization.
\end{abstract}

\maketitle

\section{Introduction}

\begin{figure}[t!]
    \centering
    \includegraphics[width=.9\linewidth]{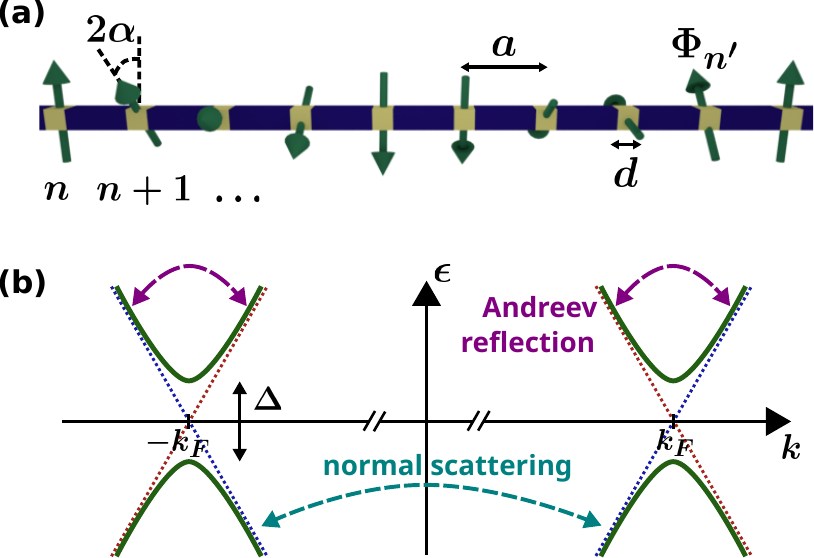}
    \caption{\textbf{(a)} Sketch of a quasi-1D helical Andreev crystal formed by a superconducting wire interrupted by magnetic regions of width $d$ separated by a constant distance $a$. It is assumed that  where $k_F^{-1} \ll d \ll \xi_0$.  The magnetization of the impurities rotates an angle $2\alpha$ around the $x$ axis between subsequent impurities. \textbf{(b)} Sketch of the superconducting spectrum with the two electron-hole (e-h)  valleys at $\pm k_F$. The semiclassical impurities forming the Andreev crystal cause only small momentum transfer processes that couple quasiparticles within the same e-h valley via Andreev scattering. Because there is no normal reflection coupling quasiparticles from opposite valleys the system presents a twofold degeneracy.}
    \label{fig:sketch}
\end{figure}

Magnetic defects and regions in a superconductor may lead to bound states that strongly change the local  spectrum\cite{yu-1965-bound, shiba-1968-classical, rusinov-1968-superconductivity, andreev-1966-electron, sakurai-1970-comments, yazdani-1997-probing, balatsky-2006-impurity, franke-2011-competition, meng-2015-superconducting, heinrich-2018-single, farinacci-2018-tuning, rouco-2019-spectral}. When the magnetic exchange coupling is larger than  the Fermi energy and the size of the magnetic region is small compared to the superconducting coherence length, a pair of non-degenerate states with opposite energies appear inside the superconducting gap. These are the so-called   Yu-Shiba-Rusinov (YSR) states\cite{yu-1965-bound,shiba-1968-classical,rusinov-1968-superconductivity}. In contrast,  if the exchange coupling is small compared to the Fermi energy, $\mu$, a pair of degenerate bound states appear\cite{andreev-1966-electron, konschelle-2016-ballistic, rouco-2019-spectral}. The origin of such degeneracy can be understood from a semiclassical perspective: electrons at the Fermi level  traveling through the
magnetic region are not back-scattered,  but they accumulate a phase, $\Phi$.  This phase  has the opposite sign for electrons/holes and spin up/down.  This results in double-degenerate bound states formed by electrons from the Fermi valleys at either $+k_F$ or $-k_F$ (see Fig.~\ref{fig:sketch}b). In a normal
metal, the phase accumulated can be gauged out. In contrast, if the host material is a superconductor,   the Andreev reflection at the semiclassical impurities leads to the coupling between electrons and holes at the same Fermi valley. This mechanism leads to Andreev bound states inside the superconducting gap. The crossover from the YSR to  Andreev limit has been studied in detail  in Ref. \cite{rouco-2019-spectral}.

In a periodic arrangement of magnetic impurities, as for example a chain, the single-impurity bound states hybridize and form bands within the superconducting gap. 
Such bands have been widely studied for atomic-sized magnetic impurities\cite{nadj-perge-2013-proposal,pientka-2013-topological,heimes-2014-majorana,poyhonen-2014-majorana,weststrom-2015-topological,pientka-2015-topological,brydon-2015-topological,schecter-2016-self,hoffman-2016-topological}. The hybridization of YSR states can lead to topological phases which host Majorana bound states at the ends of the impurity chain. 
In Ref.~\cite{rouco-2021-gap}  we proposed the analog of such atomic chains in a mesoscopic structure with lateral dimensions smaller than the superconducting coherence length, $\xi_0$.  The magnetic impurities  are replaced by  semiclassical magnetic regions realized, for example, by the contact to magnetic materials (see the sketch in Fig.~\ref{fig:sketch}a).  Mesoscopic structures involving superconductors and ferromagnetic materials  have been extensively studied, mainly in the diffusive limit, in the context of superconducting spintronics\cite{buzdin2005proximity,bergeret2005odd,beckmann-2004-evidence,khaire2010observation,banerjee2014evidence,singh2015colossal,linder2015superconducting,bakurskiy-2015-proximity}.   Our focus here is from a very different perspective. We consider clean  superconducting wires  with a periodic array of magnetic regions as  a mesoscopic realization of  crystals, which we call \textit{Andreev crystals} (ACs).


Specifically, in this article we present the general theory of  ACs, including non-collinear magnetization orientation and arbitrary separation between the magnetic impurities.  In a first part  we consider chains of impurities with non-collinear magnetization where the magnetic regions are separated by a distance $a\gtrsim \xi_0$. 
We solve the nearest-neighbours tight-binding equations of helical ACs, where the exchange field rotates a constant angle $2\alpha$ around a fixed axis between subsequent magnetic impurities, whereas their strength remains constant, $\Phi$.
The spectrum of helical ACs for energies within the superconducting gap, $|\epsilon| < |\Delta|$, shows a pair of \textit{Andreev} bands with symmetric energy with respect to the Fermi level.
In ferromagnetic configurations ($\sin\alpha=0$) the Andreev bands cross without interacting, closing the gap in a finite range of $\Phi$ values around half-integer values of $\Phi/\pi$.
Otherwise, the Andreev bands touch each other only at half-integer values of $\Phi/\pi$ forming a Dirac point.
In junctions between semi-infinite helical ACs where the rotation, $\alpha$, remains constant all along the chain and the magnetic phase changes from $\Phi_L$ to $\Phi_R$ at the left and right sides of the junction, respectively, states bounded to the interface may appear when $\text{sign}(\tan\Phi_L) \neq \text{sign}(\tan\Phi_R)$.
We refer to the junctions fulfilling this condition as \textit{inverted junctions} of ACs and they maintain similarities with Dirac system with a spatial mass inversion\cite{jackiw-1976-solitons, su-1979-solitons, su-1980-solitonb, volkov-1985-two}.
We show that inverted junctions of (anti)ferromagnetic ACs (always) never support interfacial states and that the range of parameters, $\Phi_{L(R)}$, for which the bound states appear increases as the rotation approaches an antiferromagnetic ordering (\textit{i.e.}, with decreasing value of $|\cos\alpha|$).

In a second part, we present exact calculations of the spectral properties of (anti)ferromagnetic ACs and junctions beyond the tight-binding approximation used in previous works \cite{rouco-2021-gap}. 
Specifically, we solve the  Eilenberger equation and obtain the quasiclassical Green's functions (GFs) in different situations.  
The magnetic regions are described by effective boundary conditions that take into account the spin-dependent jump of the phase, $\Phi$.
On the one hand, our solution provides the exact energy and spatial distribution of the density of states and spin polarization of the system.  
On the other hand, our results demonstrate the validity  of the first neighbours' tight-binding  approximation regarding  the gap closing in infinite antiferromagnetic ACs at half-integer values of $\Phi / \pi$, the appearance of a pair of states bounded to the interface between two antiferromagnetic ACs with inverted gaps,  and  the fractionalization of the surface spin polarization in  such junctions\cite{rouco-2021-gap}. 
 
This article is organized as follows: in Sec.~\ref{sec:model} we introduce the model Hamiltonian and the main equations used for the nearest-neighbours tight-binding model (Sec.~\ref{sec:tight-binding}),  and the Eilenberger GFs  (Sec.~\ref{sec:eilenberger}). 
In Sec.~\ref{sec:helical-ACs} we solve the nearest-neighbours tight-binding equations of infinite and junctions between semi-infinite helical ACs,  \textit{i.e.}, ACs where the exchange field of the semiclassical impurities form an helix along the wire.
In Sec.~\ref{sec:collinear-ACs} we focus on ACs where the exchange field of all the impurities is collinear. 
In particular, we solve the Eilenberger equation to obtain the quasiclassical GF of ACs with magnetic impurities following (Sec.~\ref{sec:eilenberger-ferro-acs}) ferromagnetic and (Sec.~\ref{sec:eilenberger-antiferro-acs}) antiferromagnetic ordering.
In Sec.~\ref{sec:eilenberger-junctions} we present the method to solve the Eilenberger equation in junctions between semi-infinite collinear ACs and we apply it to obtain the quasiclassical GFs in junctions between anti-ferromagnetic ACs.
Finally, in Sec.~\ref{sec:conclusions} we summarize the main results of the paper.

\section{The Model and main equations}
\label{sec:model}

We consider  a superconducting wire of lateral dimensions much smaller than the superconducting coherence length,  $\xi_0$. The wire  contains  magnetic regions
located at the points $X_n = na$, where $a$ is the separation between the
impurities and $n$ is the impurity index. We assume that the  width of the magnetic regions, $d$ is larger than  $k_F^{-1}$ and hence can be considered within the semiclassical approach\cite{rouco-2019-spectral}. In addition, we also assume that  $d \ll \xi_0$ such that we can treat the magnetic regions as  point-like  impurities,  in the semiclassical scale,  with a polarization strength and direction proportional to the corresponding SU(2) magnetic phase\cite{konschelle-2016-ballistic,konschelle-2016-semiclassical},
\begin{equation}
  \label{eq:magnetic-phase-definition}
  \hat{\boldsymbol{\sigma} \cdot}\boldsymbol{\Phi}_n \equiv \frac{1}{\hbar v_F} \int dx \; \hat{\boldsymbol{\sigma}} \cdot \boldsymbol{h}_n(x).
\end{equation}
Here, $v_F$ is the Fermi velocity, and  $\boldsymbol{h}_n(x)$ is the exchange field vector induced by the $n$-th impurity which is assumed to be parallel to the local magnetization of the magnetic region. 
The Bogoliubov-de Gennes (BdG) Hamiltonian\cite{gennes-1966-superconductivity} describing
the AC reads,
\begin{equation}
  \label{eq:BdG-Hamiltonian-general}
  \check H^\eta_{BdG} (x) = -i\eta\hbar v_F \hat\tau_3 \partial_x + \hat\tau_1\Delta
  - \hbar v_F \sum_n \hat{\boldsymbol{\sigma}} \cdot \boldsymbol{\Phi}_n \delta(x-X_n),
\end{equation}
where $\hat \tau_i$ are the Pauli matrices spanning the Nambu space
(\textit{i.e.} the electron-hole space), $\hat{\boldsymbol{\sigma}} \equiv
(\hat\sigma_1, \hat\sigma_2, \hat\sigma_3)$ stands for the vector of Pauli
matrices that span the spin space and $\eta = \pm$ refers to the two
electron-hole valleys at $\pm k_F$ (see Fig.~\ref{fig:sketch}b). A distinctive feature of semiclassical
impurities is that they do not trigger back-scattering processes. This allows
us to treat the two Fermi valleys separately and to drop the $\eta$ index. In the (first order) BdG equation, 
 Eq.~\eqref{eq:BdG-equations}, the delta functions describe the boundary conditions within  the semiclassical approach. Namely,  they describe the phase gained by a quasiparticle when it traverses the magnetic region [see Eq.~\eqref{eq:boundary-conditions-tight-binding} below].

The solution of  the BdG equations   provides  all the   spectral information about the crystal. 
As it will be shown in Sec.~\ref{sec:tight-binding}, one can solve this problem analytically under  the assumption that magnetic impurities are weakly coupled to each other, $e^{-a/\xi_0} \ll 1$. In this limit the system can be described by an  effective tight-binding model which provide the spectrum of this system.  A drawback of this approach is that to compute observable quantities, such as the local density of states or the local spin density, one has to  perform explicit summation over the Bloch momentum. 
Indeed, for calculation of observables it is more convenient to use the quasiclassical  Eilenberger equation\cite{eilenberger-1968-transformation}. 
This formalism  is presented in  Sec.~\ref{sec:eilenberger}.  
Specifically, we show how to  obtain exact analytical expressions for the quasiclassical Green's functions (GFs) of periodic ACs, and how to access to observables in a rather simple way. 
Thus, both formalims presented in Secs. ~\ref{sec:tight-binding}-Sec.~\ref{sec:eilenberger} are complementary and provides a full description of ACs.

\subsection{Tight-binding equations}
\label{sec:tight-binding}
To obtain  the spectral properties of an  AC one needs to solve the  BdG equations,
\begin{equation}
  \label{eq:BdG-equations}
  \check H_{BdG}(x) \check \Psi(x) = \epsilon \check \Psi(x),
\end{equation}
where $\check H_{BdG}(x)$ is the Hamiltonian,
Eq.~\eqref{eq:BdG-Hamiltonian-general}, and $\check \Psi(x)$ is a 4-component spinor in the Nambu$\times$spin space. The general solution of
Eq.~\eqref{eq:BdG-equations} in the region between two neighbouring impurities,
$X_n < x < X_{n+1}$, reads
\begin{equation}
  \label{eq:general-solution-BdG}
  \check \Psi (x) = {B}^+_{n+1} e^{\frac{x-X_{n+1}}{\xi}}\ket{+} +
  {B}^-_n e^{-\frac{x-X_n}{\xi}} \ket{-}.
\end{equation}
Here $\xi \equiv \frac{\hbar v_F}{\sqrt{\Delta^2 - \epsilon^2}}$ is the
superconducting coherence length, $B_n^{+(-)}$ is a 2-component spinor
(covering the spin space) that contains the amplitudes of the contributions to
the wavefunction that decays from the $n$-th impurity into the left (right), and
\begin{equation}
  \label{eq:basis-pm-ket}
  \ket{\pm} \equiv \frac{e^{\pm i\theta/2}}{\sqrt{2\cos\theta}} \left(
    \begin{array}{c}
      1 \\ \pm ie^{\mp i\theta}
    \end{array}\right),
\end{equation}
are 2-component spinors in the Nambu space, where $e^{i\theta} \equiv
\frac{\sqrt{\Delta^2-\epsilon^2} + i\epsilon}{\Delta}$ is the Andreev
factor. Direct product is assumed between the spinors in Nambu and spin spaces.

Within the semiclassical limit, quasiparticles travelling through the $n$-th semiclassical impurity do not
back-scatter, but pick up a phase according to:
\begin{equation}
  \label{eq:boundary-conditions-tight-binding}
  \check \Psi (X_n^R) = e^{i\hat\tau_3 \boldsymbol{\hat\sigma}\cdot \boldsymbol{\Phi}_n} \check \Psi(X_n^L).
\end{equation}
because of  $\hat \tau_3$ and $\boldsymbol{\hat\sigma}\cdot \boldsymbol{\Phi}_n$, the sign of the accumulated phase is different for electron/holes and spin up/down quasiparticles along the exchange field direction, respectively. Applying
these boundary conditions to the general wavefunction in
Eq.~\eqref{eq:general-solution-BdG} we obtain the equations for the $B^\pm$
coefficients, which can be recast into an effective tight-binding model by
keeping terms up to first order in $e^{-a/\xi}$. In particular, in the limit
where $e^{-a/\xi} \ll 1$, coefficients $B_n^-$ at each site $n$ can be
related to their counterparts, $B_n^+$, as follows:
\begin{equation}
  \label{eq:B-minus-to-B-plus}
  B_n^- = i \hat{\boldsymbol{\sigma}}_n
  \frac{\Delta  \sin \Phi_n}{\sqrt{\Delta^2-\epsilon^2}} B_n^+,
\end{equation}
where $\Phi_n = |\boldsymbol{\Phi}_n|$ is the strength of the magnetic phase
vector, and we define $\hat{\boldsymbol{\sigma}}_n \equiv
\frac{\hat{\boldsymbol{\sigma}} \cdot
  \boldsymbol{\Phi}_n}{\Phi_n}$. It is convenient to introduce the re-scaled coefficients, $b_n'
\equiv \hat{\boldsymbol{\sigma}}_n \sin\Phi_n B_n^+$, which satisfy a tight-binding-like equation
\begin{equation}
  \label{eq:general-tb-eigenvalue-problem}
  \Big(\omega - \hat{\boldsymbol{\sigma}}_n\omega_{0n}\Big) b_n'
  = \hat{\boldsymbol{\sigma}}_{n+1}t_{n+1} b_{n+1}' + \hat{\boldsymbol{\sigma}}_n t_n b_{n-1}'.
\end{equation}
Here $\omega \equiv \frac{\epsilon}{\sqrt{\Delta^2 - \epsilon^2}}$,  $\hat t_n \equiv -\frac{e^{-a/\xi}}{\sin\Phi_n}$ is the hopping
amplitude,  and  $\omega_{0n} =
\frac{\cos\Phi_n}{\sin\Phi_n}$ is the value of the function  $\omega$ evaluated at the bound state energy in the $n$-th  impurity, $\epsilon_{0n} = \frac{|\sin\Phi_n|}{\tan\Phi_n}$. In principle, Eq. \eqref{eq:general-tb-eigenvalue-problem} describes an arbitrary AC with lattice constant $a$.  In Ref.\cite{rouco-2021-gap} it was solved for  collinear magnetization of the impurities.   In
Sec.~\ref{sec:helical-ACs} we analyze   helical ACs composed by identical
magnetic impurities with an spatially rotating  magnetization, forming a helix in the $y$-$z$ plane.

\subsection{Eilenberger equation}
\label{sec:eilenberger}

Because of its simplicity, the tight-binding formulation, Eq.~\eqref{eq:general-tb-eigenvalue-problem},  is very useful for describing  the spectral properties of ACs. 
However, one should bear in mind that it has been derived within 
first-neighbours approximation, and therefore it is valid as long as $e^{a/\xi} \ll 1$. To go beyond this approximation we introduce here the Eilenberger equation\cite{eilenberger-1968-transformation} from which we can determine the quasiclassical  Green's functions (GFs). 

We focus again on point-like semiclassical magnetic impurities. 
The Eilenberger equation in the regions between the impurities has a simple form: 
\begin{equation}
  \label{eq:general-eilenberger-equation}
  \hbar v_F\partial_x \check g(x) - \big[i\epsilon\hat\tau_3 + \Delta \hat\tau_2, 
\; \check g(x)\big] = 0.
\end{equation}
Here $\check g(x)$ is the quasiclassical Green's function (GF), which is a 4$\times$4 matrix in the Nambu$\times$spin space that satisfies the normalization condition, $\check g^2 = 1$. The square brackets stand for the commutation operation. $\Delta$ is  the superconducting gap, which is assumed to be constant  along the superconducting wire.  Solving Eq.~\eqref{eq:general-eilenberger-equation} we obtain the propagation of the GF along the superconducting region,
\begin{equation}
  \label{eq:GF-propagation-S-region}
  \check g(x) = \hat u(x-x_0) \check g(x_0) \hat u(x_0-x),
\end{equation}
where the propagator reads,
\begin{equation}
  \label{eq:propagator-S-region}
  \hat u(x-x_0)  = \hat P_+ e^{(x-x_0)/\xi} + \hat P_- e^{-(x-x_0)/\xi}.
\end{equation}
Here $\hat P_\pm \equiv \ket{\pm} \bra{\tilde \pm} = \frac{e^{\pm i
    \hat\tau_3\theta} \pm \hat\tau_2}{2\cos\theta}$ are two orthogonal projectors
that span the Nambu space, $\ket{\pm}$ are the basis column vectors of
Eq.~\eqref{eq:basis-pm-ket} and
\begin{equation}
  \label{eq:cobasis-pm-bra}
  \bra{\tilde \pm} \equiv \frac{e^{\pm i\theta/2}}{\sqrt{2\cos\theta}} \Big(
  1 \qquad \mp ie^{\mp \theta} \Big),
\end{equation}
are the co-basis row vectors orthonormal to $\ket{\pm}$. The inverse of the propagator in Eq~\eqref{eq:propagator-S-region} fulfills the relation $[\hat u(\tilde x)]^{-1} = \hat u(-\tilde x)$.

Additionally, the GF at the right and left sides of the $n$-th semiclassical
impurity ($X_n^R$ and $X_n^L$, respectively) are connected by a propagation-like
boundary conditions,
\begin{equation}
  \label{eq:boundary-conditions-eilenberger}
  \check g(X_n^R) = e^{i\hat\tau_3\hat{\boldsymbol{\sigma}}\cdot \boldsymbol{\Phi}_n} \check g(X_n^L)
  e^{-i\hat\tau_3\hat{\boldsymbol{\sigma}}\cdot \boldsymbol{\Phi}_n}.
\end{equation}
This expression together with Eq.~\eqref{eq:propagator-S-region}, determines the GF at any space point provided its value at a given point, $\check g(x_0)$. 

In an infinite periodic ACs we need to match the
value of the GF at equivalent points of different unit cells. 
For this sake, it is useful to introduce  the \textit{chain propagator}, $\check S$, that describes the propagation of the quasiclassical GF from a given position inside a unit cell to the equivalent position in the subsequent unit cell, $\check g(x_0 + l) = \check S \check g(x_0) \check S^{-1}$ (here $l$ denotes the length of the unit cell). 
The exact form of $\check S$ depends on the arrange of impurities and the choice of the initial point inside the unit cell, $x_0$. 
Here we  choose for $x_0$ the left interface of one of the magnetic impurities. Thus, the \textit{chain propagator} reads 
\begin{equation}
    \label{eq:chain-propagator}
    \check S \equiv \prod_{j=1}^J  \hat u(a) e^{i\hat\tau_3\hat{\boldsymbol \sigma}\cdot \boldsymbol{\Phi}_j},   
\end{equation}
where $J$ is the number of impurities forming the unit cell.
The value of the quasiclassical GF at $x_0$ is obtained from the periodicity along the unit cell, $\check g(x_0) = \check S \check g(x_0) \check S^{-1}$, together with the normalization condition, $\check [g(x_0)]^2 = 1$.
Once $\check g(x_0)$ is determined the full  quasiclassical GF, $\check g(x)$, is obtained after propagation   using Eqs.~\eqref{eq:propagator-S-region} and \eqref{eq:boundary-conditions-eilenberger}.

From the knowledge  of the
GF we can obtain the local density of states (LDOS), 
\begin{equation}
  \label{eq:ldos-from-gf}
  \nu(x, \epsilon) = \text{Re} \bigg\{ \frac{1}{4} \text{Tr} \Big[ \hat\tau_3 \check g(x, \epsilon)\Big]\bigg\},
\end{equation}
and  the local spin density,
\begin{equation}
  \label{eq:local-spin-density-from-gf}
  s(x, \epsilon) = \frac{\hbar}{2} \text{Re} \bigg\{\frac{1}{4} \text{Tr} \Big[\hat\sigma_3 \hat\tau_3 \check g(x, \epsilon)\Big]\bigg\},
\end{equation}
where the traces run over the Nambu$\times$spin space. 
In Sec.~\ref{sec:collinear-ACs} we use
this approach to obtain the quasiclassical GFs of ferromagnetic and
antiferromagnetic ACs and we generalized this method  to  study junctions between different (anti-)ferromagnetic ACs.

\section{Helical Andreev Crystals}
\label{sec:helical-ACs}
In this section, we study the spectral properties of ACs with a periodic rotation of the magnetization of the magnetic impurities. 
For this sake, we use the tight-binding approach introduced in Sec. \ref{sec:tight-binding}. In particular,  we focus on an  AC consisting of identical magnetic impurities whose magnetization is in the  $y$-$z$ plane and rotates by a constant angle, $2\alpha$ around the $x$-axis\footnote{Because we do not include any spin-orbit interaction in our analysis any other planar rotation choice will give equivalent results.} (see Fig.~\ref{fig:sketch}a). 
The magnetic phase vector describing this situation  is given by
\begin{equation}
  \label{eq:helix-ac-Phi}
  \hat{\boldsymbol{\sigma}}_n \cdot \boldsymbol{\Phi}_n = \Phi e^{-i\hat\sigma_1\alpha n} \hat \sigma_3 e^{i\hat\sigma_1\alpha n},
\end{equation}

\begin{figure*}[t!]
  \centering
  \includegraphics[width=.9\linewidth]{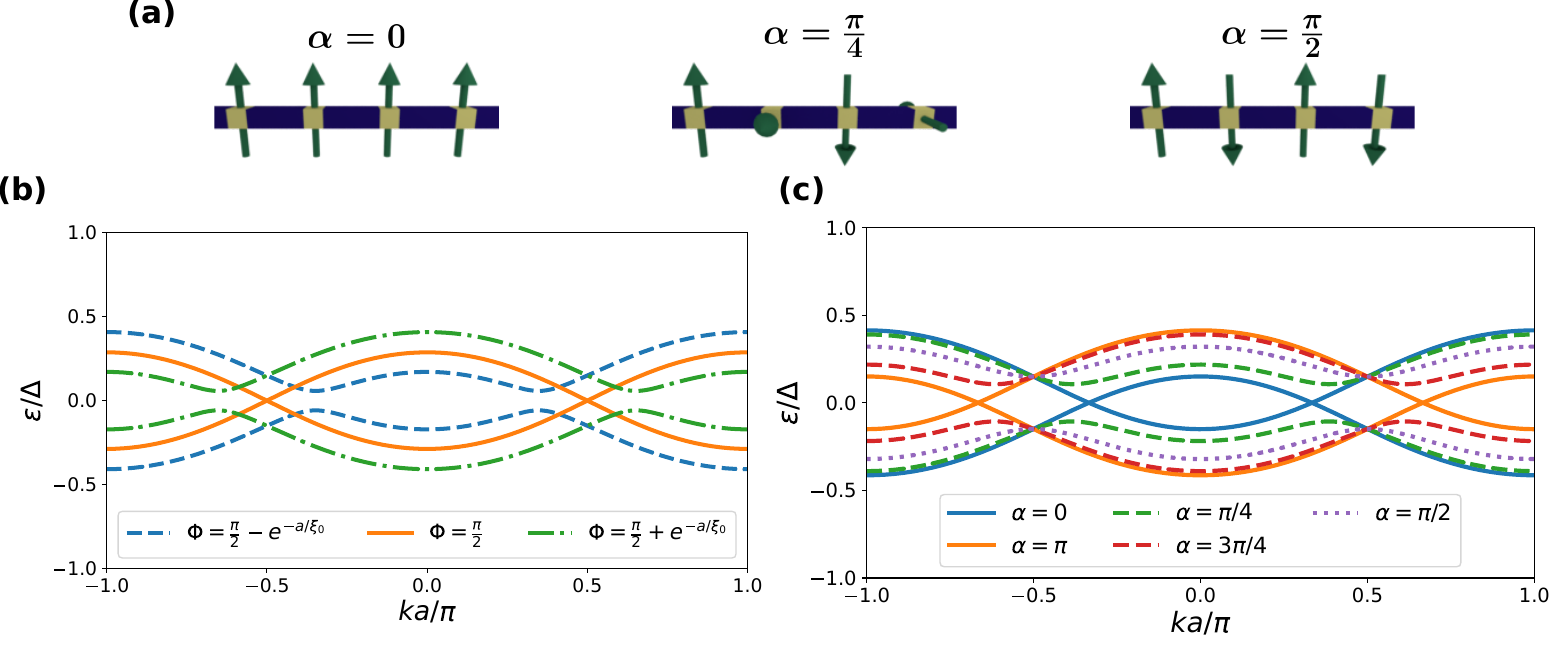}
  \caption{(a) Sketches of magnetic configurations in ACs for  different values of $\alpha$. (b) Andreev bands for  $\alpha=\frac{\pi}{4}$ and different values of $\Phi$. (c)  Andreev bands for  $\Phi=\frac{\pi}{2}-e^{-a/\xi_0}$ and
    different values of $\alpha$. 
In both panels we assumed a separation
    between impurities of $a=2\xi_0$.}
  \label{fig:helix-ac-spectrum}
\end{figure*}

where $\Phi$ is the strength of the magnetic phase. Its strength  is the same in all the
impurities. Substituting this expression into
Eq.~\eqref{eq:general-tb-eigenvalue-problem} we obtain that
\begin{equation}
  \label{eq:helix-ac-tight-binding}
  \Big[\omega - e^{i\hat\sigma_1 \alpha}\hat\sigma_3 \omega_0\Big] b_n = \hat\sigma_3 t \Big(b_{n+1} + b_{n-1}\Big),
\end{equation}
where we have defined the coefficients $b_n \equiv e^{i\hat\sigma_1 \alpha (n +
  \frac{1}{2})} b_n'$. Here  $\omega_0 = \frac{\cos\Phi}{\sin\Phi}$ stands for  the energy of the single-impurity levels and $t =
-\frac{e^{-a/\xi}}{\sin\Phi}$ is the hopping amplitude. 
After this redefinition
of the coefficients, Eq.~\eqref{eq:helix-ac-tight-binding} reduces 
to the typical tight-binding system of identical equations, whose solution  reads $b_n =
b e^{ikna}$ and $\omega = \pm\sqrt{\omega_0^2 \sin^2\alpha +
  (\omega_0\cos\alpha + 2t\cos ka)^2}$. Here $k$ is the Bloch momentum, and the spinors $b$, are obtained from
Eq.~\eqref{eq:helix-ac-tight-binding}.  The Andreev bands are defined by
\begin{equation}
  \label{eq:helix-ac-energy-bands}
  \frac{\epsilon}{\Delta} = \pm \sqrt{
    \frac{\omega_0^2 \sin^2\alpha + (\omega_0\cos\alpha + 2t\cos ka)^2}
    {1 + \omega_0^2 \sin^2\alpha + (\omega_0\cos\alpha + 2t\cos ka)^2}
  },
\end{equation}
where $t$ has to be evaluated at the energy of the single-impurity level
$\omega_0$. In Fig.~\ref{fig:helix-ac-spectrum}b and
\ref{fig:helix-ac-spectrum}c we show the subgap spectrum of ACs with different
values of $\Phi$ and $\alpha$. At $\Phi = 0$, no bound states appear, and hence
there are no Andreev bands. Increasing $\Phi$, a pair of bands emerge from the
coherent peaks and start moving towards the Fermi level, up to a point around
$\Phi = \pi/2$ where they touch each other, forming a gapless phase. Further
increase of $\Phi$ leads to a gap reopening with inverted Andreev bands.  The
latter merge with the continuum spectrum at $\Phi = \pi$. Interestingly, the
bands' inversion also happens when they merge into the continuum and reenter the
superconducting gap at $\Phi = l\pi$, where $l$ is an integer. Consequently, the
spectrum of these ACs is $\pi$ periodic in $\Phi$.

As can be seen from the energy spectrum of the bands, Eq.~\eqref{eq:helix-ac-energy-bands}, the gap closes only at half-integer values of $\Phi/2$ forming a Dirac point at $ka = \pi/2$ in ACs with any value of $\alpha$ except in those where $\sin\alpha = 0$. This situation corresponds to ferromagnetic ACs, where each of the Andreev bands corresponds to opposite spin species, and hence they do not interact while crossing. 

In Ref.~\cite{rouco-2021-gap} it was shown that a junction between two antiferromagnetic ACs with inverted gaps presents states bounded do the interface. This corresponds to the situation where $\cos \alpha = 0$ all along the structure and the sign of $\omega_0$ changes across the junction. It becomes interesting, then, to study if those bound states survive for arbitrary values of $\alpha$.
To do so, we consider a junction between two different chains where the
rotation parameter between the impurities remains constant, $\alpha$, but their
magnetic phases change from the AC on the left, $\Phi_L$, to the one on the
right $\Phi_R$. The tight-binding equations of such a system read
\begin{equation}
  \label{eq:helix-ac-junction-tb}
  \Big[\omega - e^{i\hat\sigma_1\alpha}\hat\sigma_3 \omega_{0n}\Big] b_n
  = \hat\sigma_3 t_{n+1} b_{n+1} + \hat\sigma_3 t_{n} b_{n-1},
\end{equation}
where $\omega_{0n}$ and $t_{n}$ are  defined below
Eq.~\eqref{eq:general-tb-eigenvalue-problem}. The  magnetic phase is $\Phi_L$ for $n < 0$ and $\Phi_R$ for $n \geq 0$. 
We can write for the left and right ACs,  $b_n = b_{L+}e^{-i q_{L+} n} + b_{L-}e^{-i\mu_L q_{L-} n}$ and $b_n = b_{R+} e^{i \mu_R q_{R+} n} + b_{R-} e^{i \mu_R q_{R-}
  n}$, respectively, where $q_{L(R) \pm}$ is  determined by the solution of the eigenvalue  equation, Eq.~\eqref{eq:helix-ac-junction-tb}, with positive imaginary part:
\begin{equation}
  \label{eq:helix-ac-kappa}
  \cos q_{L(R) \pm} = \frac{-\omega_{0L(R)}\cos\alpha \pm i \sqrt{\omega_{0L(R)}^2 \sin^2\alpha - \omega^2}}{2t_{L(R)}}.
\end{equation} 
According to this expression,  bound states can only appear at  energies with $\omega^2 < \omega_{0L(R)}^2\sin^2 \alpha$, \textit{i.e.}, at energies within
the gap formed by the Andreev bands of both ACs [\textit{c.f}
Eq.~\eqref{eq:helix-ac-energy-bands}].  The corresponding  eigenvectors are given by
\begin{equation}
  \label{eq:helix-ac-vectors}
  b_{L(R) \pm} = \left(
    \begin{array}{c}
      1 \\ i e^{\pm i\gamma_{L(R)}}
    \end{array}\right),
\end{equation}
where
\begin{equation}
  \label{eq:helix-ac-gamma}
e^{\pm i\gamma_{L(R)}} = \frac{- \omega \pm
  i \sqrt{\omega_{0L(R)}^2\sin^2\alpha - \omega^2}}{\omega_{0L(R)}\sin\alpha}.
\end{equation}
From  the above  results we find that bound states exist for those energies satisfying  following determinant equation:
\begin{equation}
  \left|
    \begin{array}{cccc}
      t_L & t_L & t_R & t_R \\
      t_L e^{i\gamma_L} & t_L e^{-i\gamma_L} & t_Re^{i\gamma_R} & t_R e^{-i\gamma_R} \\
      e^{i q_{L+}} & e^{i q_{L-}} & e^{-i q_{R+}} & e^{-i q_{R-}} \\
      e^{i q_{L+}} e^{i\gamma_L} & e^{i q_{L-}} e^{-i\gamma_L} & e^{-i q_{R+}}e^{i\gamma_R} & e^{-i q_{R-}} e^{-i\gamma_R} 
    \end{array}\right| = 0.
\label{eq:helix-ac-junction-determinant-equation}
\end{equation}
One can check that this equation has solutions only when $sign(\omega_{0L}) = -sign(\omega_{0R})$.  Therefore, the bound states can only appear in junctions between ACs with inverted gaps. This is a necessary but not sufficient condition. Namely,  the presence of the interfacial state in inverted junctions  depends on the magnetic rotation  along the crystal described by  $\alpha$: whereas for antiferromagnetic alignment of the impurities ($\cos\alpha = 0$) the interfacial state appears in any inverted junction, for  ferromagnetic ACs ($\sin\alpha = 0$) it never does. For any other value of $\alpha$ the existence of the bound state depends on $\Phi_L$ and $\Phi_R$ as explained below.  

\begin{figure}[t!]
  \centering
  \includegraphics[width=.9\linewidth]{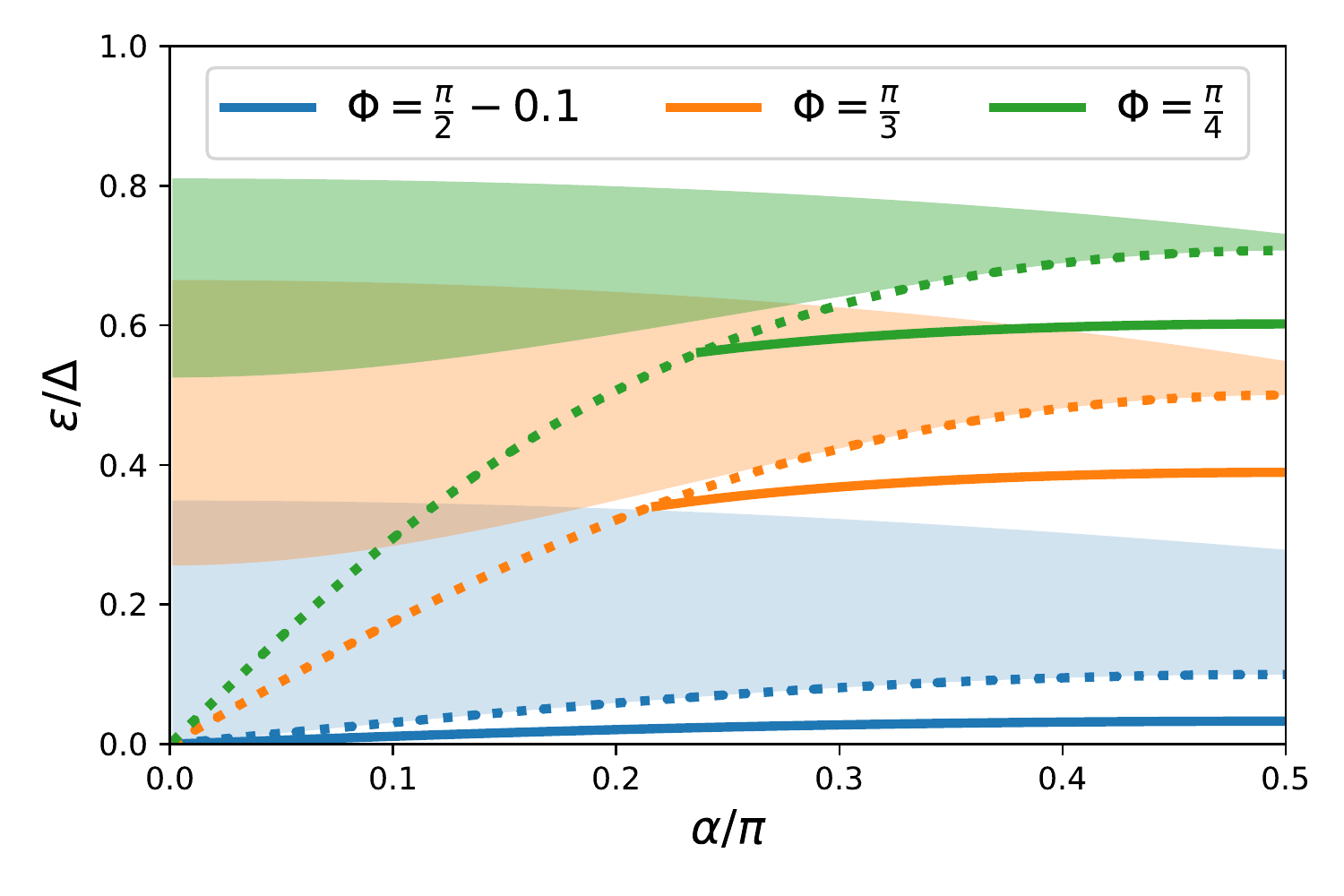}
  \caption{Energy of the (solid line) positive-energy interfacial state in terms of $\alpha$ in anti-symmetric junctions between helical ACs with $\Phi \equiv \Phi_R = -\Phi_L$. Different colors correspond to different strengths of the impurities, $\Phi$, whereas their separation all along the junction is fixed to $a=2\xi_0$. The shaded areas indicate the position of the (positive-energy) Andreev band in the respective infinite chains [Eq.~\eqref{eq:helix-ac-energy-bands}] and the dotted lines show the energy values with $\omega^2 = \omega_0^2 \sin^2\alpha$. This value determines the maximum possible energy of the bound state [Eq.~\eqref{eq:helix-ac-gamma}]. }
  \label{fig:helix-ac-junction}
\end{figure}

The determinant equation, Eq.~\eqref{eq:helix-ac-junction-determinant-equation}, can be reduced  to a compact equation in the  \textit{anti-symmetric} configuration with  $\Phi_R = -\Phi_L$. In this situation we can define  $\gamma \equiv \gamma_L = \gamma_R$ and $q_{L\pm} = q_{R\mp} \equiv \pm \kappa + i\lambda$, where $\kappa$ and $\lambda$ are real numbers determined by  Eq.\eqref{eq:helix-ac-kappa}. For $\lambda >0$, the condition for the existence of the bound state reads
\begin{equation}
    \sin^2\gamma \cosh^2\lambda - \sin^2\kappa = 0. 
    \label{eq:helix-ac-junction-sec-eq-antisymmetric}
\end{equation}

In Fig.~\ref{fig:helix-ac-junction} we show the dependence of the positive energy bound states with $\alpha$ in anti-symmetric inverted junctions of ACs with fixed value of $a = 2\xi_0$ and different strengths of the magnetic impurities, $\Phi \equiv \Phi_R = -\Phi_L$. 
With  shaded areas we show the energies within  the positive-energy  Andreev band is situated in the infinite AC [Eq.~\eqref{eq:helix-ac-energy-bands}]. 
The dotted lines correspond to energy values for which $\omega^2 = \omega_0^2\sin^2\alpha$ and they indicate the maximum possible energy of a bound state [Eq.~\eqref{eq:helix-ac-gamma}].
Close to the gap-closing point, $\cos\Phi \ll 1$, the interfacial states are present at any value of $\alpha$, excluding ferromagnetic ordering of the impurities, $\sin\alpha=0$. 
As the size of the gap between the Andreev bands increases, the range of $\alpha$ values for which the pair of bound states exist shrinks around those values corresponding to an antiferromagnetic ordering of the magnetic impurities, $\cos\alpha = 0$.

The existence or not of the bound state in anti-symmetric junctions of ACs can be understood from the relative position of the maximum-energy condition for the bound state (the dotted lines in Fig.~\ref{fig:helix-ac-junction}) and the positive-energy solution of Eq.~\eqref{eq:helix-ac-junction-sec-eq-antisymmetric}.
When $\cos\alpha \approx 0$ the maximum-energy condition locates very close to the bottom of the Andreev band and, consequently, the $\omega$ value that solves Eq.~\eqref{eq:helix-ac-junction-sec-eq-antisymmetric} almost always fulfills that $\omega^2 < \omega_0^2\sin^2\alpha$.
When $\sin\alpha \approx 0$, by contrast, the dotted lines in Fig.~\ref{fig:helix-ac-junction} approach the center of the gap, $\omega=0$.
Thus, the solution to Eq.~\eqref{eq:helix-ac-junction-sec-eq-antisymmetric} only meets the bound state existence condition, $\omega^2 < \omega_0^2\sin^2\alpha$, when the borders of the gap are also very close to the Fermi energy, \textit{i.e.}, when $\cos\Phi \approx 0$.
These considerations are also applicable in general junctions between helical ACs, in which case the energy of the bound state solves Eq.~\eqref{eq:helix-ac-junction-determinant-equation} and its existence condition is given by $\omega^2 < \text{min}(\omega^2_{0L}, \omega^2_{0R}) \sin^2\alpha$.

As a summary of  this section, for a given value of the rotation angle $\alpha$ we can classify ACs in two groups depending on  whether an interfacial state appears upon the formation of a junction between two chains with  inverted gaps.
These two groups are best exemplified by \mbox{(anti-)ferromagnetic} ACs inverted junctions in which  interfacial bound states  (always) never appear. In the next section we focus on these two type of junctions and study in more detail their spatial properties.

\section{Collinear Andreev Crystals}
\label{sec:collinear-ACs}

In this section we extend the study of \mbox{(anti-)}ferromagnetic ACs beyond the first neighbours tight-binding approximation used in previous sections. 
To do so, we solve the Eilenberger equation to obtain the quasiclassical GFs, $\check g(x)$, following the procedure discussed in Sec.~\ref{sec:eilenberger}.
From the knowledge of $\check g(x)$ we can obtain the local density of states and magnetization of ACs and junctions.

Specifically, we consider chains of magnetic impurities located at $X_n = na$, with an arbitrary separation between the impurities $a$. Here $n$ is an integer. We assume that all magnetizations, and hence the exchange fields, are aligned along the $z$ axis. Because of the collinear alignment of the exchange field we can  treat the two spin degrees of freedom separately, $\sigma = \pm$, thus reducing the size
of the GFs involved from $4 \times 4$ (in Nambu$\times$spin space) to $2 \times
2$ matrices in Nambu space. 
It follows from Eq.~\eqref{eq:GF-propagation-S-region} and the normalization condition, $[\check g(x)]^2 = 1$, that within the  region between two subsequent impurities, $X_n < x < X_{n+1}$, the quasiclassical GF for a single spin projection, $\sigma$,
can be written in terms of two independent constants, $b_{\sigma n}$ and
$c_{\sigma n}$:
\begin{align}
  \hat g_\sigma(x) = 
  &\sqrt{1 - e^{-2a/\xi} b_{\sigma n}c_{\sigma n}} \hat g^0 
  + b_{\sigma n}e^{2(x-X_{n+1})/\xi} \ket{+}\bra{\tilde -} \nonumber \\
  &+ c_{\sigma n} e^{-2(x-X_n)/\xi} \ket{-}\bra{\tilde +}.
  \label{eq:colinear-parametrized-GF}
\end{align}
Here $\hat g^{0} \equiv \hat P_+ - \hat P_-= \frac{\Delta \hat\tau_2 +
  i\epsilon\hat\tau_3}{\sqrt{\Delta^2 - \epsilon^2}}$ is the GF of an
homogeneous BCS superconductor. Equation~\eqref{eq:colinear-parametrized-GF} is the representation of the GF in the basis where the BCS propagator, Eq.~\eqref{eq:propagator-S-region}, is diagonal. 

According to Eq. (\ref{eq:boundary-conditions-eilenberger}), the GFs at the left and right side of the $n$-th impurity are 
connected by the boundary condition
\begin{equation}
  \label{eq:colinear-bc-impurity}
  \hat g_\sigma (X_n^R) = e^{i\sigma\hat\tau_3\Phi_n} \hat g_\sigma (X_n^L) e^{-i\sigma\hat\tau_3\Phi_n},
\end{equation}
where the direction to which the exchange field is pointing along the
quantization axis is determined by the sign of the magnetic phase, $\Phi_n$. Ferromagnetic ACs are described by a sequence
of identical  magnetic impurities with associated magnetic phases of $\Phi_n =
\Phi$, whereas in antiferromagnetic ACs $\Phi_n = (-1)^n \Phi$. In the next sections we study these two types of ACs and junctions between them.

\begin{figure*}[t!]
  \centering
  \includegraphics[width=\linewidth]{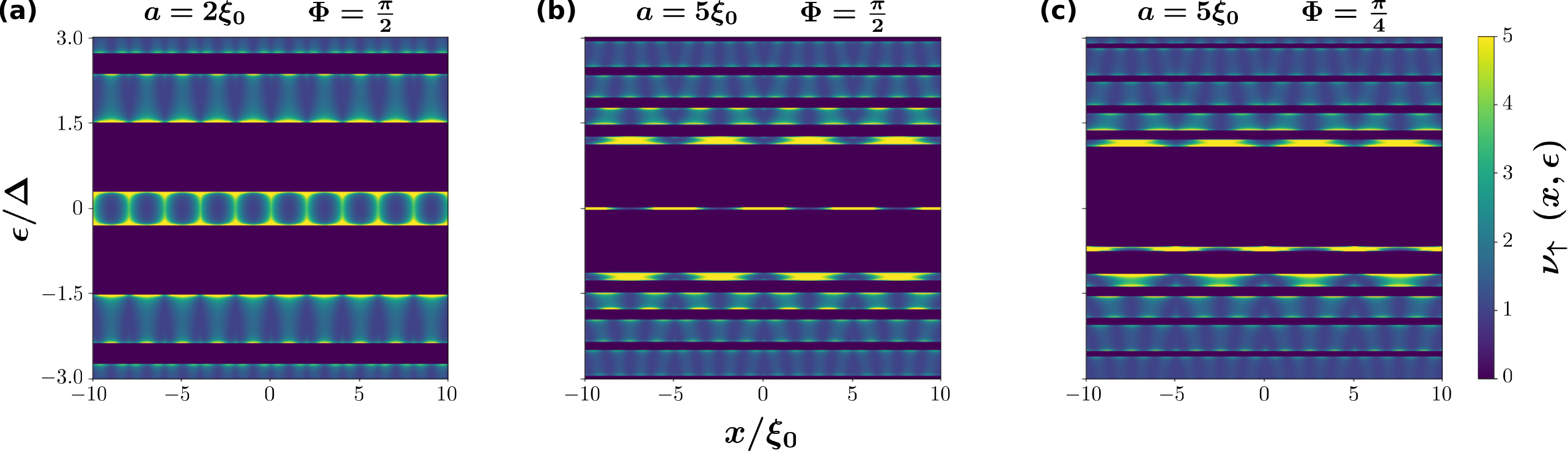}
  \caption{Local density of states, $\nu_\uparrow$, of  spin-up electrons in ferromagnetic ACs with different separation between and strengths of the magnetic impurities, $a$ and $\Phi$, respectively (see the title above each panel).}
  \label{fig:LDOS-ferro}
\end{figure*}

\subsection{Ferromagnetic ACs}
\label{sec:eilenberger-ferro-acs}

In a ferromagnetic AC the unit cell contains a single magnetic
impurity, {so the $\sigma$-spin projection of the \textit{chain propagator}, Eq~\eqref{eq:chain-propagator}, reads}
\begin{equation}
  \label{eq:eilenberger-ferro-chain-propagator}
  \hat S_{F\sigma} \equiv \hat u(a) e^{i\sigma\hat\tau_3\Phi}\; .
\end{equation}
Here $\hat u(a)$ is the BCS propagator given in  Eq.~\eqref{eq:propagator-S-region}.
The  operator $\hat S_{F\sigma}$ describes the propagation of the quasiclassical GF from the left side of  impurity $n$  to the left side  of impurity $n+1$, $\hat
g_\sigma(X_{n+1}^L) = \hat S_{F\sigma} g_\sigma(X_n^L) \hat
S_{F\sigma}^{-1}$. 

To determine the quasiclassical GF we need to obtain the  parameters $b$ and $c$  in Eq.~\eqref{eq:colinear-parametrized-GF}. The periodicity of $\hat g_\sigma$ over the unit cell, leads to 
$b_{\sigma n} = b_{\sigma}$ and $c_{\sigma n} = c_\sigma$. These expressions together with $\hat S_{F\sigma}
\hat g(X_n^L) \hat S_{F\sigma}^{-1} = \hat g(X_n^L)$ result in
\begin{equation}
  \label{eq:eilenberger-ferro-b-c-values}
  b_\sigma = c_\sigma = \frac{e^{\tfrac{a}{\xi}} \bra{\tilde +} e^{i\sigma\hat\tau_3\Phi}\ket{-}}
  {\sqrt{\left(\frac{e^{\frac{a}{\xi}} \bra{\tilde +} e^{i\sigma\hat\tau_3\Phi} \ket{+} +
        e^{-\frac{a}{\xi}} \bra{\tilde -} e^{i\sigma\hat\tau_3\Phi} \ket{-}}{2}\right)^2 - 1}}\; . 
\end{equation}
After substitution of  these values in Eq.~\eqref{eq:colinear-parametrized-GF}
one  obtains the quasiclassical GF in the magnetic regions all along  the
chain and, with it, the local density
of states (LDOS) and the local spin density [Eqs.~\eqref{eq:ldos-from-gf} and
\eqref{eq:local-spin-density-from-gf}, respectively]. 

In Fig.~\ref{fig:LDOS-ferro} we show the LDOS for a single spin species of different ferromagnetic ACs,
$\nu_\uparrow (\epsilon)$. The LDOS of
the opposite spin species can be obtained from the relation $\nu_\downarrow(\epsilon) = \nu_\uparrow(-\epsilon)$. 
The different panels in Fig.~\ref{fig:LDOS-ferro},  correspond to  different values of separation and strength of the magnetic impurities, $a$ and $\Phi$, respectively.  Within the superconducting gap, $|\epsilon| < |\Delta|$, the position of the  Andreev band depends on $\Phi$ and its width increases by decreasing $a$. 
For energies larger than $\Delta$ the continuum gets split by small gaps whose widths depend on $\Phi$, $a$ and the energy at which they lay. 
The origin of the gaps lay on the lifting of degeneracies between electronic states that differ by the reciprocal lattice vector in periodic crystals, studied in many textbooks\cite{ashcroft-1976-solid, abrikosov-2017-fundamentals, kittel-2004-introduction} At integer values of $\Phi/\pi$ the small gaps at the continuum close, whereas their width is maximum for half-integer values of $\Phi/\pi$. In the same way as it happens with the
Andreev band, the width of these small gaps increases with decreasing
$a$. The size of the gaps reduces by increasing the energy with respect to the Fermi level.

\begin{figure*}[t!]
  \centering
  \includegraphics[width=\linewidth]{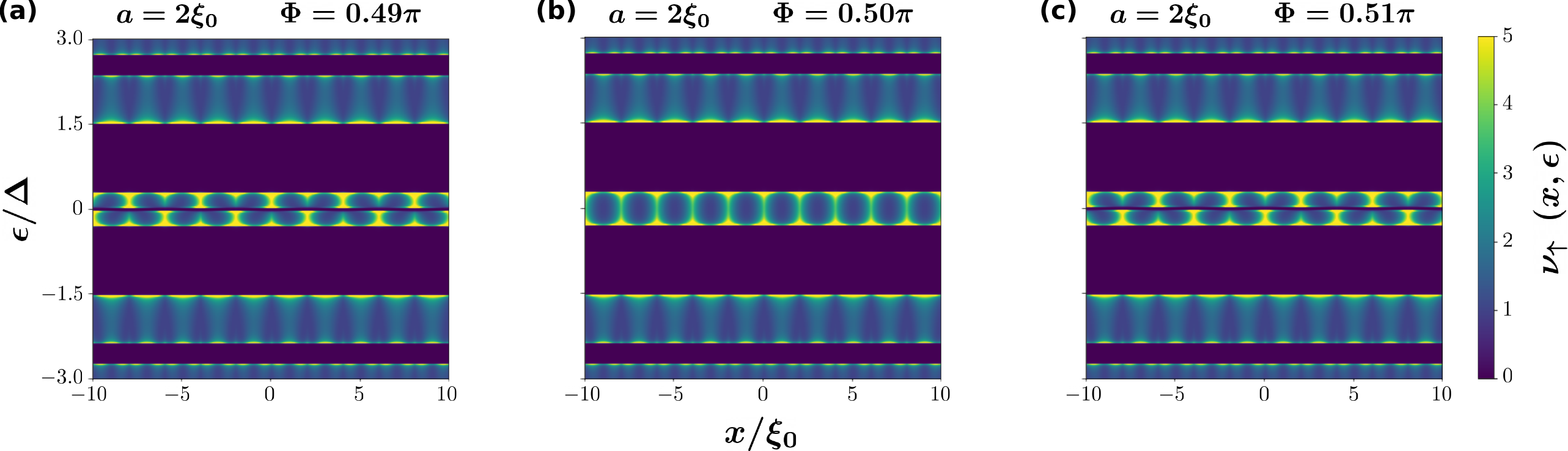}
  \caption{ Local density of states, $\nu_\uparrow$, of  spin-up electrons in an  antiferromagnetic AC close to the gap closing event, $\cos\Phi\ll 1$. The separation between impurities is  $a = 2\xi_0$.  Different panels correspond to different values of $\Phi$.}
  \label{fig:LDOS-antiferro}
\end{figure*}

\subsection{Antiferromagnetic ACs}
\label{sec:eilenberger-antiferro-acs}

In antiferromagnetic ACs the unit cell contains two identical
magnetic impurities pointing in opposite directions. The \textit{chain propagator} [Eq.~\eqref{eq:chain-propagator}] that
describes the evolution of the GF from the left interface of one magnetic
impurity to the left interface of the equivalent impurity in the next unit cell
reads,
\begin{equation}
  \label{eq:eilenberger-antiferro-chain-propagator}
  \hat S_{A\sigma} \equiv \hat u(a) e^{-i\sigma\hat\tau_3\Phi} \hat u(a) e^{i\sigma\hat\tau_3\Phi}.
\end{equation}
The unit cell consists now of  two superconducting regions with two different sets of independent parameters, namely, $b_{\sigma 0} = b_{\sigma (2n)}$, $c_{\sigma 0} = c_{\sigma(2n)}$ and $b_{\sigma 1} = b_{\sigma (2n+1)}$, $c_{\sigma 1} = c_{\sigma (2n+1)}$. 
Here $n$ is the impurity index. The  boundary condition for the impurity located between these two superconducting sections, Eq.~\eqref{eq:colinear-bc-impurity}, leads to the following relation between the set of parameters:

\begin{equation}
  \label{eq:eilenberger-antiferro-params-0-1-relation}
  b_{\sigma 1} = c_{\sigma 0}, \qquad
  c_{\sigma 1} = b_{\sigma 0}.
\end{equation}

Additionally, from the periodicity of the GF, $\hat S_{A\sigma} \hat
g_\sigma(X_{2n}^L) \hat S_{A\sigma}^{-1} = \hat g_\sigma(X_{2n}^L)$, we obtain
the expressions for
\begin{align}
  \label{eq:eilenberger-antiferro-b0-param}
  &b_{\sigma 0} = \bra{\tilde -}e^{i\sigma\hat\tau_3\Phi}\ket{-} \mathcal{D}_\sigma, \\
  \label{eq:eilenberger-antiferro-c0-param}
  &c_{\sigma 0} = -\bra{\tilde +}e^{i\sigma\hat\tau_3\Phi}\ket{+} \mathcal{D}_\sigma,
\end{align}
where
\begin{widetext}
\begin{equation}
    \label{eq:eilenberger-antiferro-D-common-factor}
    \mathcal{D}_\sigma \equiv \frac{e^{\frac{a}{\xi}} \bra{\tilde +}e^{i\sigma\hat\tau_3\Phi}\ket{-}}
    {\sqrt{\bra{\tilde +}e^{i\sigma\hat\tau_3\Phi}\ket{+}\bra{\tilde -}e^{i\sigma\hat\tau_3\Phi}\ket{-} +
    \left(\bra{\tilde +}e^{i\sigma\hat\tau_3\Phi}\ket{+}\bra{\tilde -}e^{i\sigma\hat\tau_3\Phi}\ket{-}
      \sinh \frac{a}{\xi}\right)^2}}.
\end{equation}
\end{widetext}

Substitution of these expressions  into Eq.~\eqref{eq:colinear-parametrized-GF} determines the quasiclassical  GF.  From it we obtain the LDOS for a single spin specie shown in 
Fig.~\ref{fig:LDOS-antiferro} for different values of $\Phi$ around the gap closing point, $\Phi = \pi/2$. 
The  separation between impurities is set to $a = 2\xi_0$. 
For energies within the superconducting gap a pair of Andreev bands appear at symmetric energy ranges with respect to the Fermi level. 
As it was predicted in previous calculations under the first-neighbours tight-binding approximation (Ref.~\cite{rouco-2021-gap} and Sec.~\ref{sec:helical-ACs}), these two bands touch each other only at half-integer values of $\Phi/\pi$ closing the gap around the Fermi level (see Fig.~\ref{fig:LDOS-antiferro}).
Moreover, the Andreev bands touch the continuum only when $\Phi/\pi$ is an integer: situations where the LDOS of the antiferromagnetic AC coincides with that of a pristine superconductor because the phase difference obtained by electrons and holes after propagation across an impurity is a multiple of $2\pi$.
Within  the Andreev bands, the LDOS is larger around the position of the magnetic impurities and around the energies of the single-impurity level.  
For energies larger than the  superconducting gap, $|\epsilon|>|\Delta|$,
we observe  an interference pattern and the splitting of the continuum due to the
opening of small gaps.  The  dependence of the width of the small gaps on  $\Phi$, $a$ and $\epsilon$ is the same as the one observed in ferromagnetic ACs (see the last paragraph of 
Sec.~\ref{sec:eilenberger-ferro-acs}).

\begin{figure*}[t!]
  \centering
  \includegraphics[width=\linewidth]{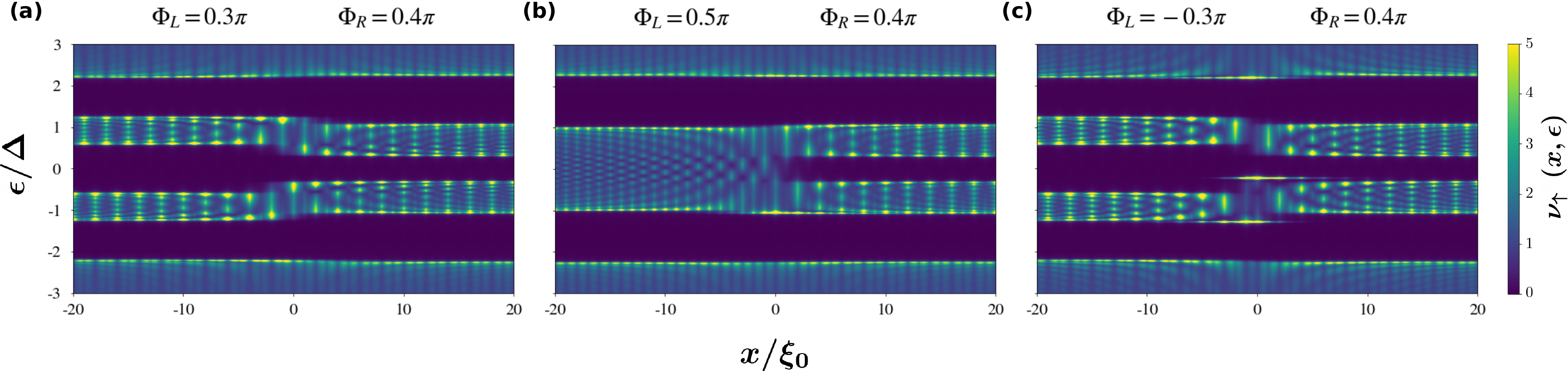}
  \caption{Local density of states  of spin-up electrons, $\nu_\uparrow(x, \epsilon)$, in  a junction between two different antiferromagnetic ACs. Inversion of the gap across the junction leads to the appearence of states bounded to the interface.}
  \label{fig:LDOS-junction}
\end{figure*}

\begin{figure}[t!]
  \centering
  \includegraphics[width=.8\linewidth]{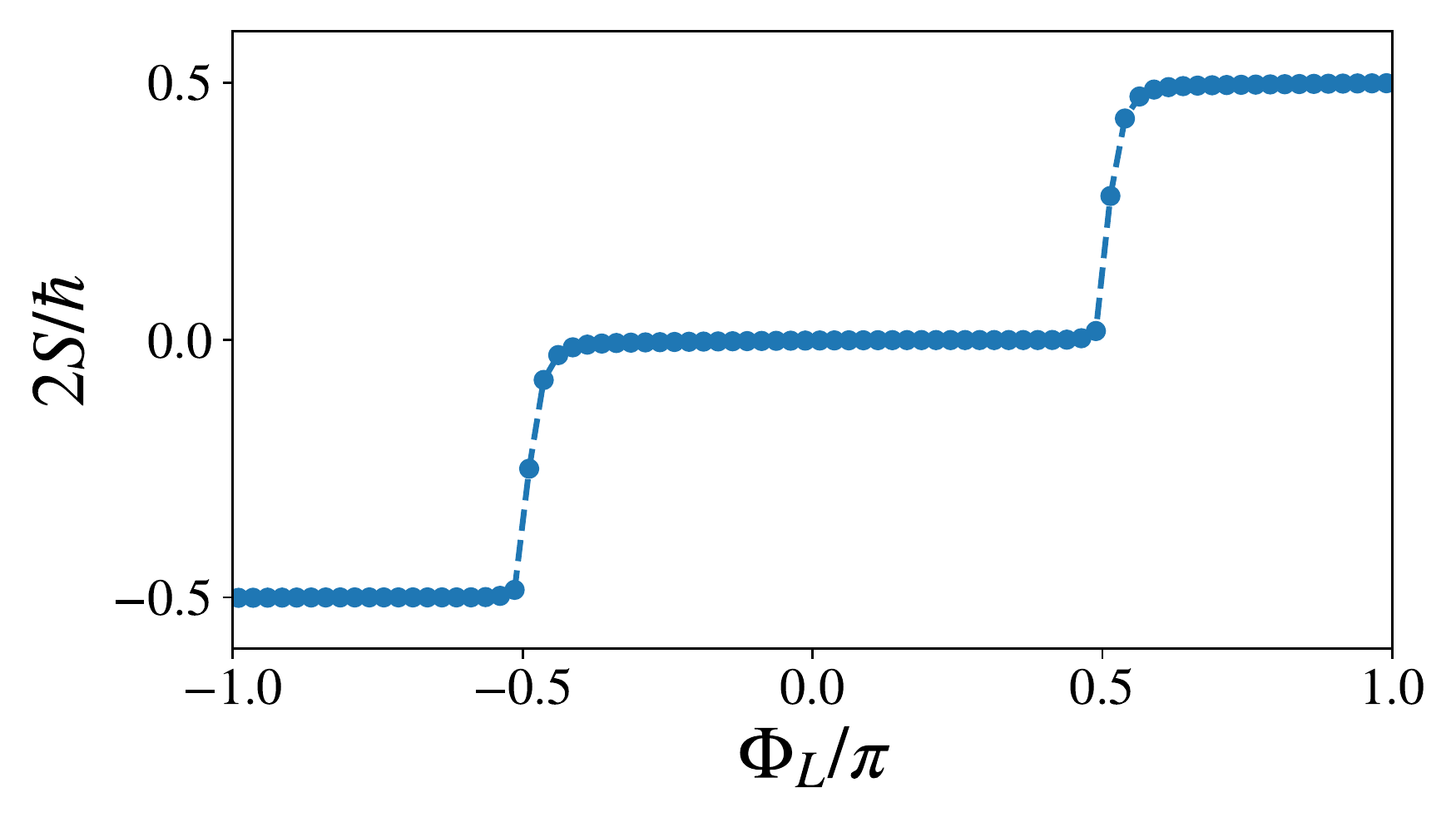}
  \caption{Contribution of a single Fermi valley to the surface spin polarization at $T=0$ of a junction between antiferromagnetic ACs in terms of $\Phi_L$ for fixed values of $\Phi_R = 0.4\pi$ and  $a=\xi_0$. The transition between plateaus is rounded due to the Dynes parameter, $\Gamma=10^{-3}\Delta$, used to avoid numerical convergence  problems.}
  \label{fig:magnetization-junction}
\end{figure}

\subsection{Junctions of collinear ACs}
\label{sec:eilenberger-junctions}
As discussed in Section~\ref{sec:helical-ACs}, inverted junctions of antiferromagnetic ACs host a pair of states bounded to the interface. 
Moreover, inverted of antiferromagnetic ACs may present fractionalization of the surface spin polarization per Fermi valley\cite{rouco-2021-gap}.
In this section we show that this result holds beyond the  tight-binding approximation used above, by solving the Eilenberger equation in junctions of ACs. 
Although we focus our analysis on junctions between antiferromagnetic ACs, the mathematical procedure presented  here is general and it can be applied to obtain the quasiclassical GFs in junctions between any type of collinear ACs.

We start by defining the $\sigma$-spin projection of the 
\textit{chain propagators} of the left(right) ACs, $\hat S_{L(R)\sigma}$, as
the operator that propagates the GFs through a  unit cell of the crystal, Eq.~\eqref{eq:chain-propagator}.
The \textit{chain propagator} is given by
Eq.~\eqref{eq:eilenberger-ferro-chain-propagator} in ferromagnetic and by
Eq.~\eqref{eq:eilenberger-antiferro-chain-propagator} in antiferromagnetic
ACs. Solving the eigenvalue problem of these operators we find a set of vectors
for which the chain propagator is diagonal,
\begin{equation}
  \label{eq:eilenberger-junctions-chain-prop-kets}
  \hat S_{L(R)\sigma} \ket{\lambda_{L(R)\sigma}^\pm} =
  e^{\pm \lambda_{L(R)\sigma}} \ket{\lambda_{L(R)\sigma}^\pm}.
\end{equation}
Because $\hat S_{L(R)\sigma}$ is, in general, not Hermitian, the
left-eigenvectors that form the co-basis
\begin{equation}
  \label{eq:eilenberger-junctions-chain-prop-bras}
  \bra{\tilde \lambda_{L(R)\sigma}^\pm} \hat S_{L(R)\sigma} =
  e^{\pm \lambda_{L(R)\sigma}} \bra{\tilde \lambda_{L(R)\sigma}^\pm},
\end{equation}
are not related by Hermitian conjugation to the right-eigenvectors in
Eq.~\eqref{eq:eilenberger-junctions-chain-prop-kets}. The eigenvectors can be
represented as exponentials with arguments of opposite sign because $\det(\hat S_{L(R)\sigma}) = 1$. 
In ferromagnetic and antiferromagnetic ACs $\lambda_\sigma$ is purely imaginary (real) for energies
where the infinite chain's spectrum shows (does not show) states.
Similarly  to the
description of the propagation  within the superconducting region
between two subsequent impurities, Eq.~\eqref{eq:GF-propagation-S-region}, the
propagation of the spin-polarized GFs through the reference points of different
unit cells reads
\begin{align}
  \hat{g}_\sigma(ml) = & \sqrt{1-v_{s\sigma} w_{s\sigma}} \Big(\ket{\lambda_{s\sigma}^+}\bra{\tilde \lambda_{s\sigma}^+} - \ket{\lambda_{s\sigma}^-}\bra{\tilde \lambda_{s\sigma}^-}\Big) \nonumber \\
  & + v_{s\sigma} e^{2\lambda_{s\sigma}m}\ket{\lambda_{s\sigma}^+}\bra{\tilde \lambda_{s\sigma}^-}
  + w_{s\sigma} e^{-2\lambda_{s\sigma}m} \ket{\lambda_{s\sigma}^-}\bra{\tilde \lambda_{s\sigma}^+}.
  \label{eq:eilenberger-junctions-propagation-GF}
\end{align}
Here $s$ is substituted by $L$ and $R$ on the left and right ACs, respectively,
$l$ is the length of the unit cell, $m$ is the unit cell index and we set the reference point inside the unit cell to $x_0 = 0$. The square root
multiplying the first term on the r.h.s. of
Eq.~\eqref{eq:eilenberger-junctions-propagation-GF} comes from the normalization
condition of the GF and the substraction of projectors that it multiplies
corresponds to the quasiclassical GF of the infinite AC at the left interface of the reference impurity.

Equation~\eqref{eq:eilenberger-junctions-propagation-GF} provides the quasiclassical GFs for a single spin, $\sigma$, at the reference points of each unit cell in terms of four parameters (two parameters per side of the junction): $v_{s\sigma}$ and $w_{s\sigma}$. 
Commensurability of $\check g(x)$ at $x \rightarrow \pm\infty$ requires that at each side of the junction one of these parameters has to be zero. Which one of the parameters is set to zero depends on the sign of $\lambda_{s\sigma}$: for $s=L$ ($s=R$) we set $w_{s\sigma}=0$ ($v_{s\sigma}=0$) when $\lambda_{s\sigma} > 0$, whereas we set $v_{s\sigma}=0$ ($w_{s\sigma}=0$) otherwise. The value of the remaining two parameters is obtained from the continuity of the quasiclassical GFs through the junction. 
Having obtained  the four parameters we  next propagate $\hat g_\sigma(ml)$ according to  Eqs.~\eqref{eq:GF-propagation-S-region} and \eqref{eq:boundary-conditions-eilenberger} to obtain the quasiclassical GFs in any position of the chain, $x$.
This method leads to analytic expressions of the quasiclassical GFs for any junction configuration.
In particular, in \ref{sec:app-af-junction} we apply this method in junctions between antiferromagnetic ACs and obtain the analytic expression of the quasiclassical GF, $\check g(x)$ [Eqs.~\eqref{eq:app-gf-final}--\eqref{eq:app-G-omega}].

In Fig.~\ref{fig:LDOS-junction} we show the obtained 
LDOS for a single spin species around the interface  between
two antiferromagnetic ACs for different values of $\Phi_L$ and fixed values of
$\Phi_R=0.4\pi$ and $a=\xi_0$. The  left panel of
Fig.~\ref{fig:LDOS-junction} shows the situation where the function of the energy of the single-impurity Andreev states, $\omega_{0}$, has the same sign at both sides of the junction. 
The spectrum exhibits a transition area around the
interface where the size of the gap between the Andreev bands changes, but no
bound states appear. 
When $\Phi_L = \pi/2$ (middle panel of Fig.~\ref{fig:LDOS-junction}) the
gap on the left side of the junction closes, whereas the gap on the right remains open.  
Further increasing of $\Phi_L$ leads to a
reopening of the left gap, as shown on the right panel of Fig.~\ref{fig:LDOS-junction}. One can clearly see how  spin-polarized bound states appear around the interface as a consequence of the gap inversion. Interestingly, these bound states are not restricted to the gap between the low-energy Andreev bands, but  appear inside all  gaps in  the spectrum.

From  the quasiclassical GF of the junction,  we can  also compute  the spin of the system by integrating
Eq.~\eqref{eq:local-spin-density-from-gf} over $x$. 
We consider the  zero-temperature case. As it is well know, quasiclassical GFs only describes the physics close to the Fermi surface and, hence, to obtain the total spin density  one has to add the Pauli paramagnetic term\cite{abrikosov-2017-fundamentals,zhang2020phase}. Namely, the Pauli paramagnetic contribution 
of each magnetic impurity is  given by \cite{rouco-2019-spectral} $\Phi/\pi$ in  units  of $\hbar / 2$. 
The resulting total value depends on the way the ACs terminate.  As we
are dealing with an infinite system, it is calculated from the average 
over all possible ending configurations of the chains\cite{rouco-2021-gap}. This is equivalent to the so-called \textit{sliding window average} method (see, for example, Sec. 4.5 of Ref.~\cite{vanderbilt-2018-berry}) and it results in a Pauli paramagnetic contribution of $\frac{\Phi_L - \Phi_R}{2\pi}$ that has to be added  to the integrated magnetization density of Eq.~\eqref{eq:local-spin-density-from-gf}. 

In Fig.~\ref{fig:magnetization-junction} we show
the contribution of a single Fermi valley to the surface spin polarization at $T=0$ of a
junction between two antiferromagnetic ACs as a function of $\Phi_L$.
We set $a=0.1\xi_0$ and $\Phi_R=0.4\pi$, although any other value of $a$ and $-\pi/2 < \Phi_R < \pi/2$ give the same results.
The magnetization per Fermi valley can only  take half-integer values of the electronic spin, which indicates fractionalization of the surface spin per electron-hole valley.  The
contribution from both Fermi valleys are equal and, hence, the total surface
magnetization equals to  an integer value of $\hbar/2$. 
Choice of $\Phi_R$ outside the range $-\pi/2 < \Phi_R < \pi/2$ would shift the ladder-like curve in Fig.~\ref{fig:magnetization-junction} some steps up or down due to the Pauli paramagnetic contribution (see previous paragraph).
In Fig.~\ref{fig:magnetization-junction} the smooth transition between plateaus is
a consequence of the small imaginary positive number that we add to the energy, $\epsilon + i\Gamma$, with  $\Gamma = 10^{-3}\Delta$, in order to avoid numerical problems. $\Gamma$ is known as Dynes parameter\cite{dynes-1978-direct} and models  the effect of inelastic scattering which leads to a  broadening of the coherent peaks peaks in the spectrum.   
In absence of inelastic processes, $\Gamma = 0$ and the magnetization shows sharp steps. 

\section{Conclusions}
\label{sec:conclusions}

In conclusion, we have presented an exhaustive study of ACs.  
We have studied the spectral properties of infinite helical ACs and junctions between them. 
For energies within the superconducting gap, the spectrum of helical ACs exhibits a pair of energy-symmetric Andreev bands with respect to the Fermi level. 
In ferromagnetic ACs ($\sin\alpha = 0$) the gap between the Andreev bands close in a finite range of $\Phi$ values around half-integer values of $\Phi/\pi$.
The range of $\Phi$ values for which the gap remains closed increasing with decreasing separation between impurities, $a$.
Otherwise, $\sin\alpha\neq 0$, the gap closes only at half-integer values of $\Phi/\pi$, forming a Dirac point.
Inverted junctions of helical ACs may present a pair of states bounded to the interface.
These states (always) never appear in inverted junctions of (anti)ferromagnetic ACs, whereas they more likely appear as the rotation of the ACs forming the inverted junction approaches an antiferromagnetic configuration (\textit{i.e.}, with decreasing value of $|\cos\alpha|$).

On the other hand, we show a method to solve the Eilenberger equation of infinite and junctions between semi-infinite ACs.
Because (anti)ferromagnetic ACs best exemplify the (existence) absence of interfacial states in inverted junctions between them, we apply this method to compute the full quasiclassical GFs of chains and junctions with collinear magnetization of the impurities. 
Our  calculations are exact and generalizes the results of Ref.~\cite{rouco-2021-gap} for arbitrary distance between the impurities, namely, that the gap around the Fermi level in antiferromagnetic ACs only closes at half-integer values of $\Phi/\pi$ and that junctions between different antiferromagnetic ACs exhibit states bounded to the interface when the gap gets inverted through the junction. 
From the quasiclassical GFs we calculate the surface spin polarization and show that such inverted junctions show fractionalization of the surface spin.  
The method that we present to solve the Eilenberger equation of collinear ACs and junctions between them can be generalized for more complex magnetic configurations. 

Overall, our results suggest the use of superconductor-ferromagnetic structures to realize crystals of a mesoscopic  scale. We predict a diversity of properties of such Andreev Crystals, as gap inversion and edge states, that can be proved by state-of-the-art spectroscopic techniques.

\section*{ Acknowledgements}
 M. R. and F.S.B.  acknowledge funding by the Spanish Ministerio de Ciencia, Innovación y Universidades (MICINN) (Project FIS2017-82804-P). 
and  EU’s Horizon 2020 research and innovation program under Grant Agreement No. 800923 (SUPERTED). I.V.T. acknowledges support by Grupos Consolidados UPV/EHU del Gobierno Vasco (Grant No. IT1249-19).

\appendix

\begin{widetext}
\section{Quasiclassical GF in a junction between antiferromagnetic ACs}
\label{sec:app-af-junction}
\newcommand{\Tp}{\braket{\tilde + | e^{i\sigma\hat\tau_3\Phi_{L(R)}} | +}}
\newcommand{\Tm}{\braket{\tilde - | e^{i\sigma\hat\tau_3\Phi_{L(R)}} | -}}
\newcommand{\Tc}{\braket{\tilde + | e^{i\sigma\hat\tau_3\Phi_{L(R)}} | -}}
\newcommand{\Tps}{\braket{\tilde + | e^{i\sigma\hat\tau_3\Phi_{s}} | +}}
\newcommand{\Tms}{\braket{\tilde - | e^{i\sigma\hat\tau_3\Phi_{s}} | -}}
\newcommand{\Tcs}{\braket{\tilde + | e^{i\sigma\hat\tau_3\Phi_{s}} | -}}
\newcommand{\af}{\frac{a}{\xi}}
\newcommand{\aaf}{\frac{2a}{\xi}}

We consider a junction between two antiferromagnetic ACs, where the separation between impurities, $a$, remains constant, but their strength changes from one chain to the other one ($\Phi_L$ and $\Phi_R$ in the left and right AC, respectively). 
Both chains meet at $x=0$. The \textit{chain propagator} of each chain is given by Eq.~\eqref{eq:eilenberger-antiferro-chain-propagator}, substituting $\Phi$ by $\Phi_L$ and $\Phi_R$ in the left and right AC, respectively. The set of eigenvalues and left- and right-eigenvectors of the chain propagator in the left (right) AC that fulfill,
\begin{equation}
  \hat S_{L(R)\sigma} \ket{\lambda_{L(R)\sigma}^\pm} = e^{\pm \lambda_{L(R)\sigma}} \ket{\lambda_{L(R)\sigma}^\pm},
  \qquad \qquad
  \bra{\tilde \lambda_{L(R)\sigma}^\pm} \hat S_{L(R)\sigma} = e^{\pm \lambda_{L(R)\sigma}} \bra{\tilde \lambda_{L(R)\sigma}^\pm},
\end{equation}
read,
\begin{align}
    &e^{\pm\lambda_{L(R)\sigma}} = 1 + 2\Tp\Tm\sinh^2 \af \pm 2 \bigg[ \Tp\nonumber \\[.5em]
    & \quad \times \Tm\sinh^2 \af + \Big(\Tp\Tm\sinh^2 \af \Big)^2\bigg]^{1/2},
    \label{eq:app-chain-propagator-eigenvalues}
\end{align}
and,
\begin{equation}
  \bra{\tilde\lambda_{L(R)\sigma}^\pm} = c_{L(R)\sigma}^\pm \bigg(\tilde d_{L(R)\sigma}^\pm \qquad 1\bigg),
  \qquad\qquad
  \ket{\lambda_{L(R)\sigma}^\pm} = c_{L(R)\sigma}^\pm \left(
    \begin{array}{c}
      1 \\ d_{L(R)\sigma}^\pm
    \end{array}\right),
    \label{eq:app-chain-propagator-eigenvectors}
\end{equation}
where,
\begin{align}
  & d_{L(R)\sigma}^\pm = \frac{e^{\pm\lambda_{L(R)\sigma}} - 1 - \big(e^{\aaf} - 1\big)\Tp\Tm}{(e^\aaf -1)\Tm\Tc}, 
  \label{eq:app-d-pm}\\[.5em]
  & \tilde d_{L(R)\sigma}^\pm = \frac{e^{\pm\lambda_{L(R)\sigma}} - 1 + \big(1 - e^{-\aaf}\big)\Tp\Tm}{(e^\aaf -1)\Tm\Tc}, 
  \label{eq:app-dt-pm}\\[.5em]
  & c_{L(R)\sigma}^\pm = \sqrt{\pm \frac{\big(e^{\aaf} - 1\big) \Tm\Tc}{2\sinh\lambda_{L(R)\sigma}}}.
  \label{eq:app-c-pm}
\end{align}
Here $\xi = \frac{\hbar v_F}{\sqrt{\Delta^2 - \epsilon^2}}$ is the superconducting coherence length. Note that $\tilde d^\pm_{L(R)\sigma} = d^\mp_{L(R)\sigma}$.  

We can parametrize the value of the quasiclassical GF at the equivalent points of the chain in terms of the eigenvectors of the \textit{chain propagator}, Eq.~\eqref{eq:app-chain-propagator-eigenvectors}, as follows:
\begin{align}
  \hat{g}_\sigma(X_{2m}^L) = &\sqrt{1-v_{s\sigma} w_{s\sigma}} \Big(\ket{\lambda_{s\sigma}^+}\bra{\tilde \lambda_{s\sigma}^+} - \ket{\lambda_{s\sigma}^-}\bra{\tilde \lambda_{s\sigma}^-}\Big) \nonumber
  \\
  & + v_{s\sigma} e^{2\lambda_{s\sigma}m}\ket{\lambda_{s\sigma}^+}\bra{\tilde \lambda_{s\sigma}^-}
  + w_{s\sigma} e^{-2\lambda_{s\sigma}m} \ket{\lambda_{s\sigma}^-}\bra{\tilde \lambda_{s\sigma}^+}.
  \label{eq:app-gf-param}
\end{align}
Here, $m$ is the unit cell index, $X_{2m}^L$ stands for the left interface of the magnetic impurity located at $X_{2m} = 2ma$ and the sub-index $s$ label the left (L) and right (R) crystal. The unit cells forming the left and right AC are those labeled by $n \leq 0$  and $m > 0$, respectively. 

For energies at which $|e^{\pm \lambda_{L(R)\sigma}}| = 1$, Eq.~\eqref{eq:app-gf-param} describes modes that propagate all along the structure. 
Otherwise, it describes exponentially decaying states by setting either $v_{L(R)\sigma}$ or $w_{L(R)\sigma}$ to zero to ensure commensurability of $\hat g_\sigma$ at the infinities. 
Which one is set to zero depends on whether $|e^{\pm \lambda_{L(R)\sigma}}| > 1$ or $|e^{\pm \lambda_{L(R)\sigma}}| < 1$. 
Indeed, numerical analysis of Eq.~\eqref{eq:app-chain-propagator-eigenvalues} shows that $|e^{\pm \lambda_{L(R)\sigma}}| \leq 1$ and, therefore, we can set $v_{L\sigma}=0$ and $w_{R\sigma}=0$.  
To obtain the remaining two parameters we require continuity of Eq.~\eqref{eq:app-gf-param} across the junction, which yields
\begin{equation}
  \label{eq:app-w-values}
  w_{L\sigma} = 2i \frac{d_{R\sigma}^+ - d_{L\sigma}^+}{d_{R\sigma}^+ - d_{L\sigma}^-},
  \qquad\qquad\qquad
  v_{R\sigma} = 2i \frac{d_{R\sigma}^- - d_{L\sigma}^-}{d_{R\sigma}^+ - d_{L\sigma}^-}.
\end{equation}
Here $d_{L(R)\sigma}^\pm$ is given by Eq.~\eqref{eq:app-d-pm}. 

Substituting Eq.~\eqref{eq:app-w-values} into Eq.~\eqref{eq:app-gf-param} we get the value of the quasiclassical GF at the left interface of every second magnetic impurity, $X_{2n}^L$.
To obtain $\hat g_\sigma(x)$ at every point inside the unit cell, hence, we have to propagate it from $X_{2n}^L$ to $x$ by means of the BCS propagator, Eq.~\eqref{eq:GF-propagation-S-region}, when the propagation is across the superconducting regions, and the propagation-like boundary conditions, \eqref{eq:boundary-conditions-eilenberger}, to connect the GFs at the left and right interfaces of each impurity.
Such a propagation allows us writing the quasiclassical GF all along the space as follows:
\begin{equation}
    \label{eq:app-gf-final}
    \hat g_\sigma(x) = 
    \begin{cases}
    \mathcal{B}_{m\sigma}^0\Big(\ket{-}\bra{\tilde -} - \ket{+}\bra{\tilde +}\Big) 
    + \mathcal{B}_{m\sigma}^{+-} e^{2\frac{x - X_{2m-1}}{\xi}} \ket{+}\bra{\tilde -} 
    + \mathcal{B}_{m\sigma}^{-+} e^{-2\frac{x - X_{2m-1}}{\xi}}\ket{-}\bra{\tilde +} \\
    \qquad\qquad\qquad\qquad\qquad\qquad\qquad\qquad\qquad\qquad 
    \text{if } X_{2n-2} < x < X_{2n-1},
    \\[2em]
    \mathcal{A}_{m\sigma}^0\Big(\ket{-}\bra{\tilde -} - \ket{+}\bra{\tilde +}\Big) 
    + \mathcal{A}_{m\sigma}^{+-} e^{2\frac{x - X_{2m}}{\xi}}\ket{+}\bra{\tilde -} 
    + \mathcal{A}_{m\sigma}^{-+} e^{-2\frac{x - X_{2m}}{\xi}}\ket{-}\bra{\tilde +}\\
    \qquad\qquad\qquad\qquad\qquad\qquad\qquad\qquad\qquad\qquad
    \text{if } X_{2n-1} < x < X_{2n},
    \end{cases},
\end{equation}
where $\ket{\pm}$ and $\bra{\tilde \pm}$ are the right- and left-eigenvectors of the BCS propagator given by Eqs.~\eqref{eq:basis-pm-ket} and \eqref{eq:cobasis-pm-bra}, respectively.
The expressions of the remaining constants depend on the side of the juction.
The $\mathcal{A}$ constants in the AC on the left ($m \leq 0$) read:
\begin{align}
    \label{eq:app-F-alpha}
    & \mathcal{A}_{m\sigma}^0 = (c_{L\sigma}^+)^2 \Big[d_{L\sigma}^+ + d_{L\sigma}^- + iw_{L\sigma} e^{-2\lambda_{L\sigma}m} d_{L\sigma}^-\Big],  \\
    & \mathcal{A}_{m\sigma}^{+-} = (c_{L\sigma}^+)^2 \Big[2 + iw_{L\sigma} e^{-2\lambda_{L\sigma}n}\Big],  \\
    & \mathcal{A}_{m\sigma}^{-+} = - (c_{L\sigma}^+)^2 d_{L\sigma}^-  
    \Big[2d_{L\sigma}^+ + iw_{L\sigma} e^{-2\lambda_{L\sigma}m} d^-_{L\sigma}\Big], 
\end{align}
whereas in the right chain ($m > 0$) they read:
\begin{align}
    & \mathcal{A}_{m\sigma}^0 = (c_{R\sigma}^+)^2 \Big[d_{R\sigma}^+ + d_{R\sigma}^- + iv_{R\sigma} e^{2\lambda_{R\sigma}m} d_{R\sigma}^+\Big],  \\
    & \mathcal{A}_{m\sigma}^{+-} = (c_{R\sigma}^+)^2 \Big[2 + iv_{R\sigma} e^{2\lambda_{R\sigma}n}\Big],  \\
    & \mathcal{A}_{m\sigma}^{-+} = - (c_{R\sigma}^+)^2 d_{R\sigma}^+  
    \Big[2d_{R\sigma}^- + iv_{R\sigma} e^{2\lambda_{R\sigma}m} d^+_{R\sigma}\Big].
    \label{eq:app-F-omega}
\end{align}
The remaining expressions for the $\mathcal{B}$-s are given in terms of the $\mathcal{A}$-s shown in Eqs.~\eqref{eq:app-F-alpha}--\eqref{eq:app-F-omega} and read
\begin{align}
  & \mathcal{B}_{m\sigma}^0 = \Big(1 + 2\Tcs\Big) \mathcal{A}_{m\sigma}^0 
  + e^{-\aaf}\Tps\Tcs \mathcal{A}_{m\sigma}^{+-} \nonumber \\
  &\qquad\qquad
  - e^{\aaf}\Tms\Tcs \mathcal{A}_{m\sigma}^{-+}, \\[1em]
  & \mathcal{B}_{m\sigma}^{+-} = 2\Tps\Tcs \mathcal{A}_{m\sigma}^0 
  + e^{-\aaf}\Tps^2 \mathcal{A}_{m\sigma}^{+-} \nonumber \\
  & \qquad\qquad
  - e^{\aaf}\Tcs^2 \mathcal{A}_{m\sigma}^{-+}, \\[1em]
  & \mathcal{B}_{m\sigma}^{-+} = -2\Tms\Tcs \mathcal{A}_{m\sigma}^0 
  - e^{-\aaf} \Tcs^2 \mathcal{A}_{m\sigma}^{+-} \nonumber \\
  & \qquad\qquad
  + e^{\aaf}\Tms^2 \mathcal{A}_{m\sigma}^{-+},
  \label{eq:app-G-omega}
\end{align}
where $\Phi_s = \Phi_L$ when $m\leq 0$ (\textit{i.e.}, in the left side of the junction) and $\Phi_s = \Phi_R$ otherwise. Equations~\eqref{eq:app-gf-final}--\eqref{eq:app-G-omega} provide the quasiclassical GF for the $\sigma$ spin component of a junction between two antiferromagnetic ACs at any position, $x$, from which we can directly calculate observables like the local density of states, Eq.~\eqref{eq:ldos-from-gf}, or the local spin density, Eq.~\eqref{eq:local-spin-density-from-gf}. 
\end{widetext}

\bibliographystyle{unsrt}
\bibliography{biblist}

\begin{thebibliography}{10}

\bibitem{yu-1965-bound}
Luh Yu.
\newblock Bound state in superconductors with paramagnetic impurities.
\newblock {\em Acta Phys. Sin}, 21(1):75, 1965.

\bibitem{shiba-1968-classical}
Hiroyuki Shiba.
\newblock Classical spins in superconductors.
\newblock {\em Progress of theoretical Physics}, 40(3):435--451, 1968.

\bibitem{rusinov-1968-superconductivity}
AI~Rusinov.
\newblock Superconductivity near a paramagnetic impurity.
\newblock {\em Zh. Eksp. Teor. Fiz. Pisma Red.}, 9:146, 1968.
\newblock [Sov. Phys. JETP {\bf 9}, 85 (1969)].

\bibitem{andreev-1966-electron}
AF~Andreev.
\newblock Electron spectrum of the intermediate state of superconductors.
\newblock {\em Zh. Eksp. Teor. Fiz.}, 49(655), 1966.
\newblock [Sov. Phys. JETP {\bf 22}, 455 (1966)].

\bibitem{sakurai-1970-comments}
Akio Sakurai.
\newblock {Comments on Superconductors with Magnetic Impurities}.
\newblock {\em Progress of Theoretical Physics}, 44(6):1472--1476, 12 1970.

\bibitem{yazdani-1997-probing}
A.~Yazdani.
\newblock Probing the local effects of magnetic impurities on
  superconductivity.
\newblock {\em Science}, 275(5307):1767–1770, Mar 1997.

\bibitem{balatsky-2006-impurity}
A.~V. Balatsky, I.~Vekhter, and Jian-Xin Zhu.
\newblock Impurity-induced states in conventional and unconventional
  superconductors.
\newblock {\em Rev. Mod. Phys.}, 78:373--433, May 2006.

\bibitem{franke-2011-competition}
K.~J. Franke, G.~Schulze, and J.~I. Pascual.
\newblock Competition of superconducting phenomena and kondo screening at the
  nanoscale.
\newblock {\em Science}, 332(6032):940–944, May 2011.

\bibitem{meng-2015-superconducting}
Tobias Meng, Jelena Klinovaja, Silas Hoffman, Pascal Simon, and Daniel Loss.
\newblock Superconducting gap renormalization around two magnetic impurities:
  From shiba to andreev bound states.
\newblock {\em Phys. Rev. B}, 92:064503, Aug 2015.

\bibitem{heinrich-2018-single}
Benjamin~W. Heinrich, Jose~I. Pascual, and Katharina~J. Franke.
\newblock Single magnetic adsorbates on s-wave superconductors.
\newblock {\em Progress in Surface Science}, 93(1):1 -- 19, 2018.

\bibitem{farinacci-2018-tuning}
La{\"e}titia Farinacci, Gelavizh Ahmadi, Ga{\"e}l Reecht, Michael Ruby, Nils
  Bogdanoff, Olof Peters, Benjamin~W. Heinrich, Felix von Oppen, and
  Katharina~J. Franke.
\newblock Tuning the coupling of an individual magnetic impurity to a
  superconductor: quantum phase transition and transport.
\newblock Jul 2018.

\bibitem{rouco-2019-spectral}
M.~Rouco, I.~V. Tokatly, and F.~S. Bergeret.
\newblock Spectral properties and quantum phase transitions in superconducting
  junctions with a ferromagnetic link.
\newblock {\em Phys. Rev. B}, 99:094514, Mar 2019.

\bibitem{konschelle-2016-ballistic}
François Konschelle, Ilya~V. Tokatly, and F.~Sebastian Bergeret.
\newblock Ballistic josephson junctions in the presence of generic spin
  dependent fields.
\newblock {\em Physical Review B}, 94(1), Jul 2016.

\bibitem{nadj-perge-2013-proposal}
S.~Nadj-Perge, I.~K. Drozdov, B.~A. Bernevig, and Ali Yazdani.
\newblock Proposal for realizing majorana fermions in chains of magnetic atoms
  on a superconductor.
\newblock {\em Phys. Rev. B}, 88(2):020407(R), jul 2013.

\bibitem{pientka-2013-topological}
Falko Pientka, Leonid~I. Glazman, and Felix von Oppen.
\newblock Topological superconducting phase in helical shiba chains.
\newblock {\em Phys. Rev. B}, 88:155420, Oct 2013.

\bibitem{heimes-2014-majorana}
Andreas Heimes, Panagiotis Kotetes, and Gerd Sch\"on.
\newblock Majorana fermions from shiba states in an antiferromagnetic chain on
  top of a superconductor.
\newblock {\em Physical Review B}, 90(6):060507(R), Aug 2014.

\bibitem{poyhonen-2014-majorana}
Kim P\"oyh\"onen, Alex Weststr\"om, Joel R\"ontynen, and Teemu Ojanen.
\newblock Majorana states in helical shiba chains and ladders.
\newblock {\em Physical Review B}, 89(11), Mar 2014.

\bibitem{weststrom-2015-topological}
Alex Weststr\"om, Kim P\"oyh\"onen, and Teemu Ojanen.
\newblock Topological properties of helical shiba chains with general impurity
  strength and hybridization.
\newblock {\em Physical Review B}, 91(6), Feb 2015.

\bibitem{pientka-2015-topological}
Falko Pientka, Yang Peng, Leonid Glazman, and Felix~von Oppen.
\newblock Topological superconducting phase and majorana bound states in shiba
  chains.
\newblock {\em Physica Scripta}, T164:014008, Aug 2015.

\bibitem{brydon-2015-topological}
P.~M.~R. Brydon, S.~Das~Sarma, Hoi-Yin Hui, and Jay~D. Sau.
\newblock Topological yu-shiba-rusinov chain from spin-orbit coupling.
\newblock {\em Physical Review B}, 91(6), Feb 2015.

\bibitem{schecter-2016-self}
Michael Schecter, Karsten Flensberg, Morten~H. Christensen, Brian~M. Andersen,
  and Jens Paaske.
\newblock Self-organized topological superconductivity in a yu-shiba-rusinov
  chain.
\newblock {\em Phys. Rev. B}, 93(14):140503 (R), Apr 2016.

\bibitem{hoffman-2016-topological}
Silas Hoffman, Jelena Klinovaja, and Daniel Loss.
\newblock Topological phases of inhomogeneous superconductivity.
\newblock {\em Phys. Rev. B}, 93:165418, Apr 2016.

\bibitem{rouco-2021-gap}
Mikel Rouco, F.~Sebastian Bergeret, and Ilya~V. Tokatly.
\newblock Gap inversion in quasi-one-dimensional andreev crystals.
\newblock {\em Physical Review B}, 103(6), Feb 2021.

\bibitem{buzdin2005proximity}
Alexandre~I Buzdin.
\newblock Proximity effects in superconductor-ferromagnet heterostructures.
\newblock {\em Reviews of modern physics}, 77(3):935, 2005.

\bibitem{bergeret2005odd}
FS~Bergeret, Anatoly~F Volkov, and Konstantin~B Efetov.
\newblock Odd triplet superconductivity and related phenomena in
  superconductor-ferromagnet structures.
\newblock {\em Reviews of modern physics}, 77(4):1321, 2005.

\bibitem{beckmann-2004-evidence}
D.~Beckmann, H.~B. Weber, and H.~v.~L\"ohneysen.
\newblock Evidence for crossed andreev reflection in superconductor-ferromagnet
  hybrid structures.
\newblock {\em Physical Review Letters}, 93(19), Nov 2004.

\bibitem{khaire2010observation}
Trupti~S Khaire, Mazin~A Khasawneh, WP~Pratt~Jr, and Norman~O Birge.
\newblock Observation of spin-triplet superconductivity in co-based josephson
  junctions.
\newblock {\em Physical review letters}, 104(13):137002, 2010.

\bibitem{banerjee2014evidence}
Niladri Banerjee, CB~Smiet, RGJ Smits, A~Ozaeta, FS~Bergeret, MG~Blamire, and
  JWA Robinson.
\newblock Evidence for spin selectivity of triplet pairs in superconducting
  spin valves.
\newblock {\em Nature communications}, 5(1):1--6, 2014.

\bibitem{singh2015colossal}
Amrita Singh, Stefano Voltan, Kaveh Lahabi, and Jan Aarts.
\newblock Colossal proximity effect in a superconducting triplet spin valve
  based on the half-metallic ferromagnet cro 2.
\newblock {\em Physical Review X}, 5(2):021019, 2015.

\bibitem{linder2015superconducting}
Jacob Linder and Jason~WA Robinson.
\newblock Superconducting spintronics.
\newblock {\em Nature Physics}, 11(4):307--315, 2015.

\bibitem{bakurskiy-2015-proximity}
Sergey~V Bakurskiy, M~Yu Kupriyanov, Artem~Alexandrovich Baranov,
  Alexandre~Avraamovitch Golubov, Nikolay~Viktorovich Klenov, and Igor~I
  Soloviev.
\newblock Proximity effect in multilayer structures with alternating
  ferromagnetic and normal layers.
\newblock {\em JETP letters}, 102(9):586--593, 2015.

\bibitem{jackiw-1976-solitons}
R.~Jackiw and C.~Rebbi.
\newblock Solitons with fermion number \textonehalf{}.
\newblock {\em Phys. Rev. D}, 13:3398--3409, Jun 1976.

\bibitem{su-1979-solitons}
W.~P. Su, J.~R. Schrieffer, and A.~J. Heeger.
\newblock Solitons in polyacetylene.
\newblock {\em Physical Review Letters}, 42(25):1698, Jun 1979.

\bibitem{su-1980-solitonb}
W.~P. Su, J.~R. Schrieffer, and A.~J. Heeger.
\newblock Soliton excitations in polyacetylene.
\newblock {\em Physical Review B}, 22(4):2099, Aug 1980.

\bibitem{volkov-1985-two}
BA~Volkov and OA~Pankratov.
\newblock Two-dimensional massless electrons in an inverted contact.
\newblock {\em JETP Lett}, 42(49):178, 1985.

\bibitem{konschelle-2016-semiclassical}
François Konschelle, F.~Sebastián Bergeret, and Ilya~V. Tokatly.
\newblock Semiclassical quantization of spinning quasiparticles in ballistic
  josephson junctions.
\newblock {\em Physical Review Letters}, 116(23), Jun 2016.

\bibitem{gennes-1966-superconductivity}
P.~G. de~Gennes.
\newblock {\em Superconductivity of Metals and Alloys}.
\newblock Benjamin, New York, 1966.

\bibitem{eilenberger-1968-transformation}
Gert Eilenberger.
\newblock Transformation of gorkov's equation for type ii superconductors into
  transport-like equations.
\newblock {\em Zeitschrift f{\"u}r Physik A Hadrons and nuclei},
  214(2):195--213, Apr 1968.

\bibitem{Note1}
Because we do not include any spin-orbit interaction in our analysis any other
  planar rotation choice will give equivalent results.

\bibitem{ashcroft-1976-solid}
Neil Ashcroft and David Mermin.
\newblock {\em Solid state physics}.
\newblock Cengage Learning, Andover England, 1976.

\bibitem{abrikosov-2017-fundamentals}
Aleksei~Alekseevich Abrikosov.
\newblock {\em Fundamentals of the Theory of Metals}.
\newblock Dover Publications, New York, 2017.

\bibitem{kittel-2004-introduction}
Charles Kittel.
\newblock {\em Introduction to Solid State Physics}.
\newblock Wiley, 8 edition, 2004.

\bibitem{zhang2020phase}
XP~Zhang, VN~Golovach, F~Giazotto, and FS~Bergeret.
\newblock Phase-controllable nonlocal spin polarization in proximitized
  nanowires.
\newblock {\em Physical Review B}, 101(18):180502, 2020.

\bibitem{vanderbilt-2018-berry}
David Vanderbilt.
\newblock {\em Berry Phases in Electronic Structure Theory}.
\newblock Cambridge University Press, Oct 2018.

\bibitem{dynes-1978-direct}
R.~C. Dynes, V.~Narayanamurti, and J.~P. Garno.
\newblock Direct measurement of quasiparticle-lifetime broadening in a
  strong-coupled superconductor.
\newblock {\em Phys. Rev. Lett.}, 41:1509--1512, Nov 1978.

\end{thebibliography}

\end{document}